\documentclass[a4paper,12pt]{article}
\usepackage{jheppub} % for details on the use of the package, please
\usepackage{tcilatex}
\usepackage{amsfonts}
\usepackage{amssymb}
\usepackage{multicol}
\usepackage{graphicx}
\usepackage{float}
\usepackage{caption}
\usepackage{xcolor}
\usepackage[utf8]{inputenc}
\usepackage{amsmath}
\title{\boldmath Finiteness of 3D higher spin gravity Landscape }
\author[1,2]{R. Sammani}
\author[1,2]{Y. Boujakhrout}
\author[1,2]{E.H Saidi}
\author[1,2]{R. Ahl Laamara}
\author[1,2]{L.B Drissi}
\affiliation[1]{ LPHE-MS, Science Faculty, Mohammed V University in Rabat, Morocco}
\affiliation[2]{Centre of Physics and Mathematics, CPM- Morocco}
\emailAdd{rajae\_sammani@um5.ac.ma}
\abstract{ We give Swampland constraints on the three dimensional Landscape of Anti-de Sitter higher spin gravity in the Chern-Simons formulation\emph{\ }with connection valued in various split real forms of Lie algebras. We derive the finiteness conjecture by computing the upper bound on the rank of possible gauge groups then we refine it using the AdS distance conjecture. We discuss the implications of this Swampland constraint on the spectrum of higher spin gravity theories and we compare it with the gravitational exclusion principle required from BTZ black hole consideration to excerpt a constraint on the Chern-Simons level k.}

\arxivnumber{2304.01887}
\begin{document}
\notoc
\maketitle
\flushbottom
\newpage
\tableofcontents
\newpage
\section{Introduction}

\label{sec:intro} Mapping bulk quantum gravitational theories to non
gravitational edge quantum field theories is undoubtedly one of the most
important breakthrough in modern physics. The AdS/CFT correspondence \cite%
{A1} provides the perfect setting to study quantum gravity since the
antecedent model $AdS_{3}/CFT_{2}$ \cite{A2,C1,C11,C2,C2A}. The lack of
local degrees of freedom in $AdS_{3}$ gravity does not effect its dynamical
properties that are predominantly determined by the choice of the boundary
conditions \cite{A33,A333,A3333} like those proposed by Brown and Henneaux
\cite{A2} and generalized recently by Grumiller and Riegler \cite{A3}. The
correlation between $AdS_{3}$ gravity and the $CFT_{2}$ theory was first
established when the 3D Einstein's general relativity was recast as a
topological field theory with holographic dynamics dictated by the 2D
boundary CFT. It was Achucarro, Townsend \cite{C1} and later on Witten \cite%
{C2} who discovered that in $AdS_{3}$ space, the action and the equations of
motion are equivalent to a Chern-Simons (CS) theory with an appropriate
gauge group. Now, we can formulate Einstein's $AdS_{3}$ gravity as a
Chern-Simons theory with a much simplified structure.

The framework of the $AdS_{3}/CFT_{2}$ model turned out to be very useful
for Black hole physics \cite{BL2}-\cite{BL6}. For instance, the study of the
BTZ black hole solution \cite{A4,A5} is of immense importance to inquire
more about black holes characteristics which makes one ponder about its
successfulness for other physics areas such as higher spin (HS) theories
\cite{HS1}-\cite{A9}. Indeed, the Chern-Simons theory based on $sl(N,\mathbb{%
R})$ algebra which is a formulation of higher spin gravity, a version of
Vasiliev's higher spin theory, played a major role in the asymptotic
analysis particularly the derivation of the $AdS_{3}$ asymptotic symmetry
algebra (ASA) \cite{A10,A11,A111}.%

In this work, we exploit a particular set of boundary conditions
that we extend from Grumiller and Riegler's (GR) boundary conditions \cite%
{A3} in the Chern-Simons formulation with connection valued in various split
real forms of Lie algebras. The relaxation of the GR constraints lead to an
enlarged asymptotic symmetry that includes an additional current algebra in
the affine boundary. In fact, one of the main purposes of the boundary
conditions is the computation of the asymptotic symmetry algebra as a way to
establish the correspondence between the bulk AdS$_{3}$\textrm{\ and the
boundary CFT}$_{2}.$\textrm{ The derivation of the asymptotic symmetries, as
will be demonstrated subsequently, is mainly based on the determination of
the corresponding boundary charge via its variation as follows: }$\delta
Q_{\zeta }=\frac{k}{2\pi }\int d\xi ^{-}tr\left[ \zeta \left( \delta
\mathfrak{A}_{-}+\delta \mathfrak{A}_{+}\right) \right] ,$ \textrm{where} $%
\mathfrak{A}_{\pm }$\ \textrm{designate the boundary field components and} $%
\zeta $\textrm{\ an infinitesimal gauge parameter. By imposing }$\delta
\mathfrak{A}_{+}=0,$\textrm{\ according to the GR boundary conditions, we
exclude a substantial component of the boundary charge which affects the
resulting asymptotic algebra. The term }$\delta \mathfrak{A}_{+}$\textrm{\
was forgone to restore the non-invariance and cancel the gauge anomaly, but
what if we can reinstate a healthy variational principle without
relinquishing }$\delta \mathfrak{A}_{+}?$\textrm{\newline
We show that this can be indeed achieved via the anomaly inflow mechanism
that allows us to explore the consequences of having boundary anomalies
while insuring their cancelation. Additionally, the relaxation of the GR
boundary conditions combined with the inflow mechanism grant us a
descriptive constraint that characterises the Swampland program finiteness
conjecture in 3D allowing us therefore to study the higher spin
gravitational theories from the lens of the Swampland program.}%

The goal of the Swampland program \cite{A13,A12222} is to define the frontier between
Landscape and Swampland theories. Basically, we are treating two types of
models: effective field theories that remain consistent when coupled to
gravity, also known as Landscape models. And effective field theories that
appear consistent but are, actually not, referred to as Swampland models. classifying a theory as either part of the Landscape or the Swampland requires
testing the theory using an ensemble of criteria termed Swampland
conjectures. It is indeed a very challenging task to determine such
criteria, and proving them, universally, is even more difficult; this is why
they are labelled as conjectures. They could be derived from many physics
areas, like black hole physics or string theory, but they should be
independent of a specific UV completion. The Swampland criteria must hold
for all low energy effective field theories that could be completed into
high energy theories regardless of the specifics of a particular UV theory.%

Our Landscape analysis in the $AdS_{3}/CFT_{2}$ framework aims to
investigate the finiteness conjecture \cite{A13,A12222,A1222,A122,A12}; stating
that a consistent effective field theory coupled to gravity must have a
finite Landscape \cite{A12222,rev}. In order to check this, we first define
an $AdS_{3}$ Landscape based on a series of higher spin families by
considering the Chern-Simons formulation of $AdS_{3}$ gravity with various
gauge symmetries, then we derive the upper bound on the corresponding rank
by studying the anomalies of the dual conformal field theory. Next, we
inspect the implications of the constraint on higher spin theories and
juxtapose it with the requirements of the AdS distance conjecture to further
sharpen it. We also provide additional support from the Literature; as we
will see, the upper bound of the rank can lead to the gravitational
exclusion principle \cite{G2} with an appropriate constraint on the level of
the corresponding Chern-Simons theory \textrm{k}.

The organisation of this paper is as follows: \textrm{In \autoref{sec2}}, we
use the CS description of AdS$_{3}$ to derive the extended\textrm{\ }%
asymptotic symmetry algebra for the $sl(2,\mathbb{R})\times sl(2,\mathbb{R})$
case. \textrm{In \autoref{sec3}}, we couple the theory to higher spin fields
by generalising the $sl(2,\mathbb{R})$ theory to higher dimensional gauge
groups which lead to a series of HS families. \textrm{\autoref{sec4}} is
fully reserved to the Landscape analysis; we start by defining the AdS$_{3}$
Landscape based on the series of higher spin families presented in section
2. Then, we compute the boundary anomaly polynomial and discuss its
cancellation. Next, we derive the Swampland finiteness constraint, evaluate
the implications on the spectrum of HS theories and study the entanglement
with the AdS distance conjecture. Finally, we compare the obtained bound
with the gravitational exclusion principle and excerpt a constraint on the
CS level $\mathrm{k}.$ \textrm{In \autoref{sec5}} we conclude our work with
a summary and a few comments. \textrm{In \autoref{app}} we give some
illustrations of the graphical implementation of the principal embedding
with Tits-Satake graphs.

\section{Extended asymptotic dynamics of AdS$_{3}$ gravity}

\label{sec2} In this section, we first describe the GR boundary conditions
of AdS$_{3}$ gravity in the CS formulation with connection valued in $sl(2,%
\mathbb{R})_{L}\times sl(2,\mathbb{R})_{R}$; these conditions were first
obtained by Grumiller and Riegler \cite{A3} lead to an affine invariance of
the AdS$_{3}$ boundary; they are given by eqs(\ref{IBCA}-\ref{IBD}) and
termed below as\ GR conditions. Then, we give an extension of these boundary
conditions by relaxing (\ref{IBD}) into (\ref{EBD}); thus enlarging the
asymptotic symmetry by including an additional current algebra.

\textrm{To begin, let us briefly review the fundamental aspects of the CS
formulation of AdS}$_{3}.$ According to Achucarro, Townsend and Witten \cite%
{C1,C2}, Einstein's gravity in 3D with a negative cosmological constant $%
\Lambda =-1/l_{AdS}^{2}$ can be remarkably expressed as a difference of two
Chern-Simons (CS) actions as follows:%
\begin{equation}
\mathcal{S}_{{\small 3D}}^{{\small grav}}\left( \mathbf{e},\mathbf{\omega }%
\right) =\mathcal{S}_{{\small CS}}^{{\small gauge}}\left( \mathbf{A}\right) -%
\mathcal{S}_{{\small CS}}^{{\small gauge}}(\mathbf{\tilde{A}).}  \label{CS}
\end{equation}%
On the left hand side of this relationship, the fields $\mathbf{e}$ and $%
\mathbf{\omega }$ are the 3D gravity gauge fields; they describe
respectively the dreibein and the spin connection. Their field action $%
\mathcal{S}_{{\small 3D}}^{{\small grav}}$ reads like:
\begin{equation}
\mathcal{S}_{{\small 3D}}^{{\small grav}}=\frac{1}{8\pi G_{{\small N}}}%
\int\nolimits_{\mathcal{M}_{3D}}tr\left( \mathbf{e}\wedge \mathbf{R}+\frac{%
\Lambda }{3!}\mathbf{e}^{3}\right) ,  \label{cs1}
\end{equation}%
with Riemann 2-form $\mathbf{R}$ given by the EOM $\delta \mathcal{S}_{%
{\small 3D}}^{{\small grav}}/\delta \mathbf{\omega }=0$ as $d\mathbf{\omega }%
+\frac{1}{2}\mathbf{\omega }\wedge \mathbf{\omega }=-\frac{\Lambda }{2}%
\mathbf{e}\wedge \mathbf{e}$. The $G_{N}^{(3D)}$ is the Newton constant in
3D; it will be set below as $G_{N}^{(3D)}=G_{N}$. Using the following
homomorphism for the AdS$_{3}$ isometry:%
\begin{equation}
SO\left( 2,2\right) \simeq SL(2)_{L}\times SL(2)_{R}.  \label{ssp}
\end{equation}%
The right hand of eq(\ref{CS}) concerns the CS gauge field action with $%
SL\left( 2,\mathbb{R}\right) _{L}\times $ $SL\left( 2,\mathbb{R}\right) _{R}$
gauge symmetry and gauge potentials $\mathbf{A},\mathbf{\tilde{A}}$; it is
given by \textrm{\cite{Ysra,usra}:}
\begin{eqnarray}
\mathcal{S}_{{\small CS}}^{{\small gauge}}\left( \mathbf{A}\right)  &=&\frac{%
k}{4\pi }\int\nolimits_{\mathcal{M}_{3D}}tr(\mathbf{Ad\mathbf{A}+}\frac{2}{3}%
\mathbf{A}^{3}),  \label{csl} \\
\mathcal{S}_{{\small CS}}^{{\small gauge}}\left( \mathbf{\tilde{A}}\right)
&=&\frac{\tilde{k}}{4\pi }\int\nolimits_{\mathcal{M}_{3D}}tr(\mathbf{\tilde{A%
}d\tilde{A}+}\frac{2}{3}\mathbf{\tilde{A}}^{3}),  \label{csr}
\end{eqnarray}%
where we have set $\mathbf{A}=\left( \mathbf{A}\right) _{L}$ and $\mathbf{%
\tilde{A}}=(\mathbf{\tilde{A})}_{R}$. The corresponding EOM gives the
following gauge field 2-forms $\mathbf{F}=d\mathbf{A}+\mathbf{A}\wedge
\mathbf{A=0},$ and $\mathbf{\tilde{F}}=d\mathbf{\tilde{A}}+\mathbf{\tilde{A}}%
\wedge \mathbf{\tilde{A}=0}$ ; they are therefore flat and accordingly
topological. The $k$ and $\tilde{k}$ are Chern-Simons levels (here $\tilde{k}%
=k$); these are positive integers related to the 3D Newton constant as $%
l_{AdS}/(4G_{N});$ thus leading to the relations \cite{C2}:%
\begin{equation}
k^{2}=\frac{l_{{\small AdS}}^{2}}{16G_{N}^{2}}=-\frac{1}{16\Lambda G_{N}^{2}}%
.  \label{level}
\end{equation}%
We define the following linear combinations of gauge fields valued in the
Lie algebra of $SL\left( 2,\mathbb{R}\right) ,$ the diagonal gauge symmetry
within $SL\left( 2,\mathbb{R}\right) _{L}\times $ $SL\left( 2,\mathbb{R}%
\right) _{R}:$
\begin{equation}
\begin{tabular}{lllll}
$\mathbf{\omega }$ & $=$ & $\omega ^{a}J_{a}$ & $=$ & $\frac{1}{2}\left(
A^{a}+\tilde{A}^{a}\right) J_{a},$ \\
$\frac{1}{l}\mathbf{e}$ & $=$ & $e^{a}J_{a}$ & $=$ & $\frac{1}{2}\left(
A^{a}-\tilde{A}^{a}\right) J_{a}.$%
\end{tabular}%
\end{equation}%
In terms of these objects, the gravitational and gauge field curvatures
valued in $sl\left( 2,\mathbb{R}\right) $ are given by:
\begin{equation}
\mathbf{R}=J_{a}R^{a}\qquad ,\qquad \mathbf{F}=(J_{a})_{L}F^{a}\qquad
,\qquad \mathbf{\tilde{F}}=(J_{a})_{R}\tilde{F}^{a}.
\end{equation}%
Where the three generators of the $sl\left( 2,\mathbb{R}\right) $ Lie
algebra $J_{a}=(J_{0},J_{1},J_{2})$ satisfy the commutation relations:
\begin{equation}
\left[ J_{a},J_{b}\right] =\varepsilon _{abc}J^{a}\qquad ,\qquad tr\left(
J_{a}J_{b}\right) =\frac{1}{2}\eta _{ab},  \label{F5}
\end{equation}%
with $\varepsilon _{012}=1$ and metric $\eta _{ab}=(-,+,+)$.

In this formulation, the derivation of the relationship (\ref{CS}) can be
computed smoothly; it is obtained by thinking about the CS gauge fields in
terms of linear combinations of the 3D gravity fields as follows:
\begin{equation}
\begin{tabular}{lllll}
$\mathbf{A}$ & $=$ & $A^{a}J_{a}$ & $=$ & $\left( \omega ^{a}+\frac{1}{l}%
e^{a}\right) J_{a},$ \\
$\mathbf{\tilde{A}}$ & $=$ & $\tilde{A}^{a}J_{a}$ & $=$ & $\left( \omega
^{a}-\frac{1}{l}e^{a}\right) J_{a},$%
\end{tabular}%
\end{equation}%
where we have set $l_{AdS}=l$ required by the scaling dimension. Using
properties of the trace combined with the antisymmetry property of odd
differential forms like $dx^{\mu }\wedge dx^{\nu }=-dx^{\nu }\wedge dx^{\mu }
$, one can check that (\ref{CS}) holds up to the total derivatives $tr[d(%
\mathbf{e\omega )]}$. The key relations in these calculations are given by:
\begin{equation}
\begin{tabular}{lll}
$tr\left( \mathbf{A}d\mathbf{A}-\mathbf{\tilde{A}}d\mathbf{\tilde{A}}\right)
$ & $=$ & $\frac{4}{l}tr\left( \mathbf{e}d\mathbf{\omega }\right) +d\left(
tr[\frac{2}{l}\mathbf{e\omega ]}\right) ,$ \\
$tr\left( \mathbf{A}^{3}-\mathbf{\tilde{A}}^{3}\right) $ & $=$ & $\frac{2}{l}%
tr\left( 3\mathbf{e\omega }^{2}+\frac{1}{l^{2}}\mathbf{e}^{3}\right) .$%
\end{tabular}
\label{F2}
\end{equation}

\subsection{GR boundary conditions on the CS connection}

In this subsection, we first recast the $SL\left( 2,\mathbb{R}\right) $
basis presented above (\ref{F5}) in a more suitable basis for both the
derivation of the asymptotic symmetries and the coupling of higher spin
fields. Then, we \textrm{rewrite} the GR boundary conditions of AdS$_{3}$
given in \cite{A3} in preparation for our extension.

\ \ \

$\bullet $ \emph{Recasting the }$SL\left( 2,\mathbb{R}\right) $ \emph{basis:}%
\newline
In the polar coordinate frame $x^{\mu }=\left( t,\rho ,\varphi \right) $ of
the 3D spacetime with $\rho \in \mathbb{R}_{+}$ being the radial coordinate
of the AdS$_{3}$ cylinder, the potential vectors of the gauge group $%
SL\left( 2,\mathbb{R}\right) _{L}\times SL\left( 2,\mathbb{R}\right) _{R}$
are given by the components $A_{\mu }=\left( A_{t},A_{\rho },A_{\varphi
}\right) $ and $\tilde{A}_{\mu }=(\tilde{A}_{t},\tilde{A}_{\rho },\tilde{A}%
_{\varphi }).$ These fields obey the \textrm{uncoupled} classical field
equations:%
\begin{equation}
\begin{tabular}{lllll}
$SL\left( 2,\mathbb{R}\right) _{L}$ & $:$ & $\partial _{\mu }A_{\nu
}^{a}-\partial _{\nu }A_{\mu }^{a}+\varepsilon _{abc}A_{\mu }^{b}A_{\nu }^{c}
$ & $=$ & $0,$ \\
$SL\left( 2,\mathbb{R}\right) _{R}$ & $:$ & $\partial _{\mu }\tilde{A}_{\nu
}^{a}-\partial _{\nu }\tilde{A}_{\mu }^{a}+\varepsilon _{abc}\tilde{A}_{\mu
}^{b}\tilde{A}_{\nu }^{c}$ & $=$ & $0,$%
\end{tabular}
\label{AS}
\end{equation}%
with $A_{\mu }=J_{a}A_{\mu }^{a}$. As these two sets of fields equations are
similar, we will only focus on the first one while keeping in mind that the
AdS$_{3}$ gravity has two gauge sectors: a left sector for $A_{\mu }\equiv
A_{\mu }^{L}$ and a right sector for $\tilde{A}_{\mu }\equiv A_{\mu }^{R}$.
Moreover, motivated by the AdS$_{3}$/CFT$_{2}$ correspondence, we will think
about the Lie algebras $sl\left( 2,\mathbb{R}\right) _{L}$ and $sl\left( 2,%
\mathbb{R}\right) _{R}$ of the gauge symmetry of the CS theory as
subalgebras of the conformal symmetry of CFT$_{2}$ given by \cite{A3,SA}:%
\begin{equation}
\begin{tabular}{lll}
$\left[ L_{n},L_{m}\right] $ & $=$ & $\left( n-m\right) L_{n+m}+\frac{c}{12}%
\left( n^{3}-n\right) \delta _{n+m},$ \\
$\left[ \bar{L}_{n},\bar{L}_{m}\right] $ & $=$ & $\left( n-m\right) \bar{L}%
_{n+m}+\frac{\bar{c}}{12}\left( n^{3}-n\right) \delta _{n+m},$%
\end{tabular}
\label{vir}
\end{equation}%
with labels as $n,m\in \mathbb{Z}$. These are two copies Vir$_{c}$ and Vir$_{%
\bar{c}}$ of the Virasoro algebra with $L_{n}$ and $\bar{L}_{n}$ generators,
it reduces to the finite dimensional symmetries, the $SL\left( 2,\mathbb{R}%
\right) _{L}\times SL\left( 2,\mathbb{R}\right) _{R}$ for a null anomalous
term corresponding to the labels $n,m=0,\pm $. For that, we use the
following new basis $K_{0,\pm }$ and $\bar{K}_{0,\pm }$ for the generators
of $SL\left( 2,\mathbb{R}\right) _{L}\times SL\left( 2,\mathbb{R}\right)
_{R};$ they satisfy the commutation relations:
\begin{equation}
\begin{tabular}{lllll}
$SL\left( 2,\mathbb{R}\right) _{L}$ & $:$ & $\left[ K_{n},K_{m}\right] $ & $=
$ & $\left( n-m\right) K_{n+m},$ \\
$SL\left( 2,\mathbb{R}\right) _{R}$ & $:$ & $\left[ \bar{K}_{n},\bar{K}_{m}%
\right] $ & $=$ & $\left( n-m\right) \bar{K}_{n+m},$%
\end{tabular}
\label{H4}
\end{equation}%
with labels $n,m=0,\pm $. These are subalgebras of the infinite Virasoro
symmetry (\ref{vir}). The $K_{0,\pm }$ (and similarly $\bar{K}_{0,\pm }$)
are related to the basis $J_{a}$ used before as follows:%
\begin{equation}
\begin{tabular}{lllllllllll}
$K_{+}$ & $=$ & $J_{0}+J_{1}$ & $,\qquad $ & $K_{-}$ & $=$ & $J_{0}-J_{1}$ &
$,\qquad $ & $K_{0}$ & $=$ & $J_{2},$ \\
$2J_{0}$ & $=$ & $K_{+}+K_{-}$ & $,\qquad $ & $2J_{1}$ & $=$ & $K_{+}-K_{-}$
& $,\qquad $ & $J_{2}$ & $=$ & $K_{0}.$%
\end{tabular}
\label{H3}
\end{equation}%
By denoting the triplet $\left( K_{+},K_{0},K_{-}\right) =K_{a}$ and $%
K^{a}=\eta ^{ab}K_{b}$ as just $K_{-a}$; and moreover thinking about the
commutation relation (\ref{H4}) of the new generators like $\left[
K_{a},K_{b}\right] =\Upsilon _{ab}^{a+b}K_{a+b}$, we end up with the
following expressions of $\mathrm{\Upsilon }_{ab}^{c},$ $\eta _{ab}$ and $%
\kappa _{ab}=tr\left( K_{a}K_{b}\right) :$
\begin{equation}
\mathrm{\Upsilon }_{ab}^{c}=\left( a-b\right) \delta _{a+b}^{c},\qquad \eta
_{ab}=\left(
\begin{array}{ccc}
0 & 0 & 1 \\
0 & 1 & 0 \\
1 & 0 & 0%
\end{array}%
\right) ,\qquad \kappa _{ab}=\left(
\begin{array}{ccc}
0 & 0 & -1 \\
0 & \frac{1}{2} & 0 \\
-1 & 0 & 0%
\end{array}%
\right) .  \label{cst}
\end{equation}

$\bullet $ \emph{Asymptotic behaviour and radial gauge:}\newline
The boundary of AdS$_{3}$ corresponds to the limit $\rho \rightarrow \rho
_{\infty }$ , which is a non compact surface $\Sigma _{1,1}$ that can be
imagined as an infinite cylinder given by the fibration $\mathbb{S}_{\varphi
}^{1}\times \mathbb{R}_{t}$ with compact coordinate $0\leq \varphi \leq 2\pi
$ and non compact axis given by the time-like direction $t$. On the boundary
of AdS$_{3},$ the CS gauge field $A_{\mu }^{a}$ in the bulk splits as $%
\left( \mathfrak{A}_{\rho }^{a},\mathfrak{A}_{\alpha }^{a}\right) $ with
label $\alpha =t,\varphi $. The boundary field $\mathfrak{A}_{\alpha }^{a}$
is given by the asymptotic value of the bulk $A_{\mu }^{a}$; that is:%
\begin{equation}
\begin{tabular}{lll}
$A_{\alpha }^{a}\left( t,\rho ,\varphi \right) $ & $\qquad \rightarrow
\qquad $ & $\mathfrak{A}_{\alpha }^{a}\left( \xi ^{\pm }\right) ,$ \\
$A_{\rho }^{a}\left( t,\rho ,\varphi \right) $ & $\qquad \rightarrow \qquad $
& $\mathfrak{A}_{\rho }^{a}=cte,$%
\end{tabular}
\label{BC}
\end{equation}%
valued in $sl(2,\mathbb{R}).$ In this mapping, the boundary gauge field $%
\mathfrak{A}_{\alpha }^{a}$ is a function only of the variable $\xi ^{\pm
}=(\varphi \pm t)/\sqrt{2}.$ Similarly, we have:%
\begin{equation}
\begin{tabular}{lll}
$\tilde{A}_{\alpha }^{a}\left( t,\rho ,\varphi \right) $ & $\qquad
\rightarrow \qquad $ & $\widetilde{\mathfrak{A}}_{\alpha }^{a}\left( \xi
^{\pm }\right) ,$ \\
$\tilde{A}_{\rho }^{a}\left( t,\rho ,\varphi \right) $ & $\qquad \rightarrow
\qquad $ & $\widetilde{\mathfrak{A}}_{\rho }^{a}=cte.$%
\end{tabular}
\label{CB}
\end{equation}%
This mapping can be implemented by using $\left( \mathbf{i}\right) $ the
gauge covariant derivatives $D=d+A$ in the AdS$_{3}$ bulk and the $\mathfrak{%
D}=d+\mathfrak{A}$ in the boundary $\Sigma _{1,1};$ and $\left( \mathbf{ii}%
\right) $ the $SL\left( 2,\mathbb{R}\right) _{L}\times SL\left( 2,\mathbb{R}%
\right) _{R}$ gauge symmetry allowing to relate the two 1-forms as follows:%
\begin{equation}
\begin{tabular}{lllllll}
$D$ & $=$ & $\mathfrak{g}^{-1}\mathfrak{Dg}$ & $\qquad ,\qquad $ & $%
\mathfrak{A}$ & $=$ & $\mathfrak{A}\left( \xi ^{\pm }\right) ,$ \\
$\tilde{D}$ & $=$ & $\mathfrak{g}\widetilde{\mathfrak{D}}\mathfrak{g}^{-1}$
& $\qquad ,\qquad $ & $\widetilde{\mathfrak{A}}$ & $=$ & $\widetilde{%
\mathfrak{A}}\left( \xi ^{\pm }\right) .$%
\end{tabular}
\label{H1}
\end{equation}%
Below, we focus on the left sector $D=\mathfrak{g}^{-1}\mathfrak{Dg};$ that
is for $SL\left( 2,\mathbb{R}\right) _{L}.$ In the mapping (\ref{H1}), the
asymptotic gauge field $\mathfrak{A}\left( \xi ^{\pm }\right) $ is the
boundary gauge potential with expansion like $\mathfrak{A}=\mathfrak{A}%
_{+}d\xi ^{+}+\mathfrak{A}_{-}d\xi ^{-}$. The $\mathfrak{g}=\mathfrak{g}%
\left( \rho ,t,\varphi \right) $ is a group element $\exp [\vartheta \left(
\rho ,t,\varphi \right) ]$ of the gauge symmetry $SL\left( 2,\mathbb{R}%
\right) _{L}$ taken in its Borel subgroup as follows \cite{A3}:%
\begin{equation}
\mathfrak{g}=e^{\mathrm{\beta }_{0}K_{0}}e^{\mathrm{\beta }_{\mathbf{+}}K_{%
\mathbf{-}}}\qquad ,\qquad \mathrm{\beta }_{0}=1,\qquad \mathrm{\beta }_{%
\mathbf{+}}=\rho .  \label{H2}
\end{equation}%
On the boundary there is no $\rho $- component, so we denote the two
existing component of the boundary gauge field like $\mathfrak{A}_{\alpha }=(%
\mathfrak{A}_{+},\mathfrak{A}_{-})$. Using the boundary limit (\ref{BC}),
the gauge field equation (\ref{AS}) in the bulk gets mapped into:
\begin{equation}
\partial _{+}\mathfrak{A}_{-}-\partial _{-}\mathfrak{A}_{+}+\left[ \mathfrak{%
A}_{+},\mathfrak{A}_{-}\right] =0.  \label{fe}
\end{equation}%
Moreover, using the GR expansions in \cite{A3}:
\begin{eqnarray}
\mathfrak{A}_{-} &=&\frac{-2\pi }{\mathrm{k}}\left[ \mathfrak{A}%
_{-}^{+}K_{+}-2\mathfrak{A}_{-}^{0}K_{0}+\mathfrak{A}_{-}^{-}K_{-}\right] ,
\label{IBCA} \\
\mathfrak{A}_{+} &=&\mathfrak{A}_{+}^{+}K_{+}+\mathfrak{A}_{+}^{0}K_{0}+%
\mathfrak{A}_{+}^{-}K_{-},  \label{IBCA2}
\end{eqnarray}%
with%
\begin{eqnarray}
\delta \mathfrak{A}_{-} &=&\frac{-2\pi }{\mathrm{k}}\left[ \delta \mathfrak{A%
}_{-}^{+}K_{+}-2\delta \mathfrak{A}_{-}^{0}K_{0}+\delta \mathfrak{A}%
_{-}^{-}K_{-}\right] ,  \label{IBC} \\
\delta \mathfrak{A}_{+} &=&0,  \label{IBD}
\end{eqnarray}%
the field equation of motion (\ref{fe}) splits as follows:%
\begin{equation}
\begin{tabular}{lll}
$\partial _{+}\mathfrak{A}_{-}^{0}-\frac{\mathrm{k}}{4\pi }\partial _{-}%
\mathfrak{A}_{+}^{0}+\left( \mathfrak{A}_{+}^{+}\mathfrak{A}_{-}^{-}-%
\mathfrak{A}_{+}^{-}\mathfrak{A}_{-}^{+}\right) $ & $=$ & $0,$ \\
$\partial _{+}\mathfrak{A}_{-}^{+}+\frac{\mathrm{k}}{2\pi }\partial _{-}%
\mathfrak{A}_{+}^{+}-\left( 2\mathfrak{A}_{+}^{+}\mathfrak{A}_{-}^{0}+%
\mathfrak{A}_{+}^{0}\mathfrak{A}_{-}^{+}\right) $ & $=$ & $0,$ \\
$\partial _{+}\mathfrak{A}_{-}^{-}+\frac{\mathrm{k}}{2\pi }\partial _{-}%
\mathfrak{A}_{+}^{-}+\left( 2\mathfrak{A}_{+}^{-}\mathfrak{A}_{-}^{0}+%
\mathfrak{A}_{+}^{0}\mathfrak{A}_{-}^{-}\right) $ & $=$ & $0.$%
\end{tabular}
\label{H5}
\end{equation}

\subsection{Extended GR boundary conditions}

A preliminary inspection of the GR boundary conditions, particularly the
variation of the boundary gauge field components (\ref{IBD}), makes one
wonder why only $\mathfrak{A}_{-}$ components are allowed to vary. Well,
this is due to the requirement of a good variational principle. In fact,
starting from:
\begin{equation}
\mathcal{S}_{CS}\left[ A\right] =\frac{k}{4\pi }\int_{\mathcal{M}%
_{3D}}tr(AdA+\frac{2}{3}A^{3}),
\end{equation}%
and performing an infinitesimal gauge change $A\rightarrow A+\delta A$, we
find:
\begin{equation}
\delta \mathcal{S}_{CS}\left[ A\right] =\frac{k}{2\pi }\int_{\mathcal{M}%
_{3D}}tr\left[ \delta A\left( dA+A^{2}\right) \right] +\frac{k}{4\pi }%
\int_{\partial \mathcal{M}_{3D}}tr(\delta AA),
\end{equation}%
indicating that the action has no extremum on-shell as it doesn't vanish by
using the on-shell field equation $F=dA+A^{2}=0$. At the boundary, the
1-form potential $A=A_{t}dt+A_{\varphi }d\varphi $ is expressed like $%
\mathfrak{A}_{+}d\xi ^{+}+\mathfrak{A}_{-}d\xi ^{-}$ (for short $\mathfrak{A}%
_{\alpha }d\xi ^{\alpha }$) as given by (\ref{BC}), the variation becomes:%
\begin{equation}
\begin{tabular}{lll}
$\mathfrak{A}\wedge \mathfrak{A}$ & $=$ & $\left( \mathfrak{A}_{+}\mathfrak{A%
}_{-}-\mathfrak{A}_{-}\mathfrak{A}_{+}\right) d\xi ^{+}\wedge d\xi ^{-},$ \\
$\delta \mathfrak{A}\wedge \mathfrak{A}$ & $=$ & $\left( \delta \mathfrak{A}%
_{+}\mathfrak{A}_{-}-\delta \mathfrak{A}_{-}\mathfrak{A}_{+}\right) d\xi
^{+}\wedge d\xi ^{-}.$%
\end{tabular}%
\end{equation}%
Consequently, we have:%
\begin{equation}
\delta \mathcal{S}_{CS}=\frac{k}{4\pi }\int_{\partial \mathcal{M}%
_{3D}}d^{2}\xi \sqrt{\left\vert \mathrm{h}\right\vert }\left( \delta
\mathfrak{A}_{+}\mathfrak{A}_{-}-\delta \mathfrak{A}_{-}\mathfrak{A}%
_{+}\right) ,
\end{equation}%
with $\mathrm{h}$ the determinant of the boundary metric $h_{\alpha \beta
}\left( \xi ^{+},\xi ^{-}\right) $. By adding the external term $\mathcal{S}%
_{CS}^{bnd}$ given by:%
\begin{equation}
\mathcal{S}_{CS}^{bnd}=\frac{k^{\prime }}{4\pi }\int\nolimits_{\partial
\mathcal{M}_{3D}}d^{2}\xi \sqrt{\left\vert \mathrm{h}\right\vert }tr(%
\mathfrak{A}_{+}\mathfrak{A}_{-}),
\end{equation}%
the gauge variation $\delta \mathcal{S}_{tot}^{CS}$ becomes for the case $%
k^{\prime }=k:$
\begin{equation}
\delta \mathcal{S}_{tot}^{CS}=\delta (\mathcal{S}_{CS}+\mathcal{S}%
_{CS}^{bnd})=\frac{k}{2\pi }\int\nolimits_{\partial \mathcal{M}%
_{3D}}d^{2}\xi \sqrt{\left\vert \mathrm{h}\right\vert }tr\left[ \delta
\mathfrak{A}_{+}\mathfrak{A}_{-}\right] .  \label{bndy}
\end{equation}%
A good variational principle is therefore restored by taking $\delta
\mathfrak{A}_{+}=0.$

What we suggest to do is to take $\delta \mathfrak{A}_{+}\neq 0$ instead. In
fact, the derivation of the asymptotic symmetries, as we will see in the
upcoming subsection, is mainly based on the computation of the corresponding
boundary charge via its variation as follows:%
\begin{equation}
\begin{tabular}{lll}
$\delta _{\zeta }Q$ & $=$ & $\frac{k}{2\pi }\int tr\left[ \zeta \left(
\delta \mathfrak{A}_{-}d\xi ^{-}+\delta \mathfrak{A}_{+}d\xi ^{+}\right) %
\right] ,$ \\
& $=$ & $\frac{k}{2\pi }\int tr\left( \zeta \delta \mathfrak{A}_{-}\right)
d\xi ^{-},$ \\
& $\equiv $ & $\delta _{\zeta }Q^{(-)}.$%
\end{tabular}
\label{233}
\end{equation}%
\textrm{Evidently by taking} $\delta \mathfrak{A}_{+}=0$ as in \cite{A3} we
forgo a substantial component of the boundary charge and the associated
symmetry, which affects the overall resulting asymptotic algebra. By setting
$\delta _{\zeta }Q=\delta _{\zeta }Q^{(-)}+\delta _{\zeta }Q^{(+)}$ with $%
\delta _{\zeta }Q^{(\pm )}$ as in the above relation, we see that in the GR
boundary conditions, we have $\delta _{\zeta }Q^{(+)}=0.$ Obviously, the
term $\delta \mathfrak{A}_{+}$ was relinquished to restore the
non-invariance and cancel the gauge anomaly, but what if we can reinstate a
healthy variational principle without foregoing $\delta \mathfrak{A}_{+}.$
Indeed, this can be achieved via the anomaly inflow mechanism that allows us
to explore the boundary anomalies while insuring their cancelation.
Additionally, the relaxation of the GR boundary conditions combined with the
inflow mechanism will grant us a descriptive constraint equation given by (%
\ref{ccL}) that characterises the Swampland program finiteness conjecture in
3D incorporating therefore higher spin gravity within the general quantum
gravitational \textrm{Landscape.}

For these purposes, our boundary conditions (Extended GR conditions) will
then take the form (\ref{IBCA}) with:%
\begin{eqnarray}
\delta \mathfrak{A}_{-} &=&\frac{-2\pi }{\mathrm{k}}\left[ \delta \mathfrak{A%
}_{-}^{+}K_{+}-2\delta \mathfrak{A}_{-}^{0}K_{0}+\delta \mathfrak{A}%
_{-}^{-}K_{-}\right] \neq 0,  \label{EBC} \\
\delta \mathfrak{A}_{+} &=&\frac{-2\pi }{\mathrm{k}}\left[ \delta \mathfrak{A%
}_{+}^{+}K_{+}+\delta \mathfrak{A}_{+}^{0}K_{0}+\delta \mathfrak{A}%
_{+}^{-}K_{-}\right] \neq 0.  \label{EBD}
\end{eqnarray}%
As stated before, this will induce an important repercussion in the form of
a gauge anomaly; in lieu of cancelling it the traditional way as in \cite{A3}
, we will acknowledge and investigate the consequences of this breakdown of
non-invariance and what it might entail. Moving forward, our plan of action
regarding the anomaly consists of two main steps. First, we shall explore
the implications of the unfolded gauge non-invariance on the asymptotic
symmetries; see subsection \textbf{2.3}. Then, we consider the cancellation
of this anomaly using the anomaly inflow mechanism; see subsection \textbf{%
4.2.}

\subsection{Enlarged asymptotic gauge symmetry}

This subsection will address two key points: first, we compute the
asymptotic symmetry algebra (ASA) of the GR boundary conditions following
the standard method \textrm{\cite{ASA}}. Then, we derive the generalised
current algebra associated to the boundary term (\ref{bndy}).

\paragraph{ GR asymptotic symmetry algebra:}

Here, we construct the asymptotic Lie algebra (ASA) of the gauge symmetry
pertaining the GR boundary conditions (\ref{H1}). For that, we proceed as
follows: First, we investigate the gauge transformations of the 1-form gauge
potential $\mathfrak{A}$ on the AdS$_{3}$ boundary. For later use, we
express this gauge potential in various ways as follows:%
\begin{equation}
\begin{tabular}{lll}
$\mathfrak{A}$ & $=$ & $\mathfrak{A}_{+}d\xi ^{+}+\mathfrak{A}_{-}d\xi ^{-},$
\\
$\mathfrak{A}_{\alpha }$ & $=$ & $K_{+}\mathfrak{A}_{\alpha }^{+}+K_{0}%
\mathfrak{A}_{\alpha }^{0}+K_{-}\mathfrak{A}_{\alpha }^{-}.$%
\end{tabular}
\label{a}
\end{equation}%
with label $\alpha =\pm .$ The $sl(2,\mathbb{R})$ Lie algebra expansion of $%
\mathfrak{A}_{-}$ is given by (\ref{IBCA}); and that of $\mathfrak{A}_{+}$
is as in (\ref{IBCA2}). Next, we derive the GR-ASA which turns out to be
just the affine Lie algebra $sl(2,\mathbb{R})_{\mathrm{k}}$ with Kac-Moody
(KM) level $\mathrm{k}.$

\

\textbf{A) Gauge symmetry of Eq}(\ref{a})\newline
The gauge transformation of the boundary field $\mathfrak{A}=\mathfrak{A}%
_{\alpha }d\xi ^{\alpha }$ is obtained by starting from the gauge
transformation of the bulk\ CS gauge connection $A=K_{a}A_{\mu }^{a}dx^{\mu }
$ as depicted by Figure \textbf{\ref{GT}}:
\begin{figure}[tbph]
\begin{center}
\includegraphics[width=14cm]{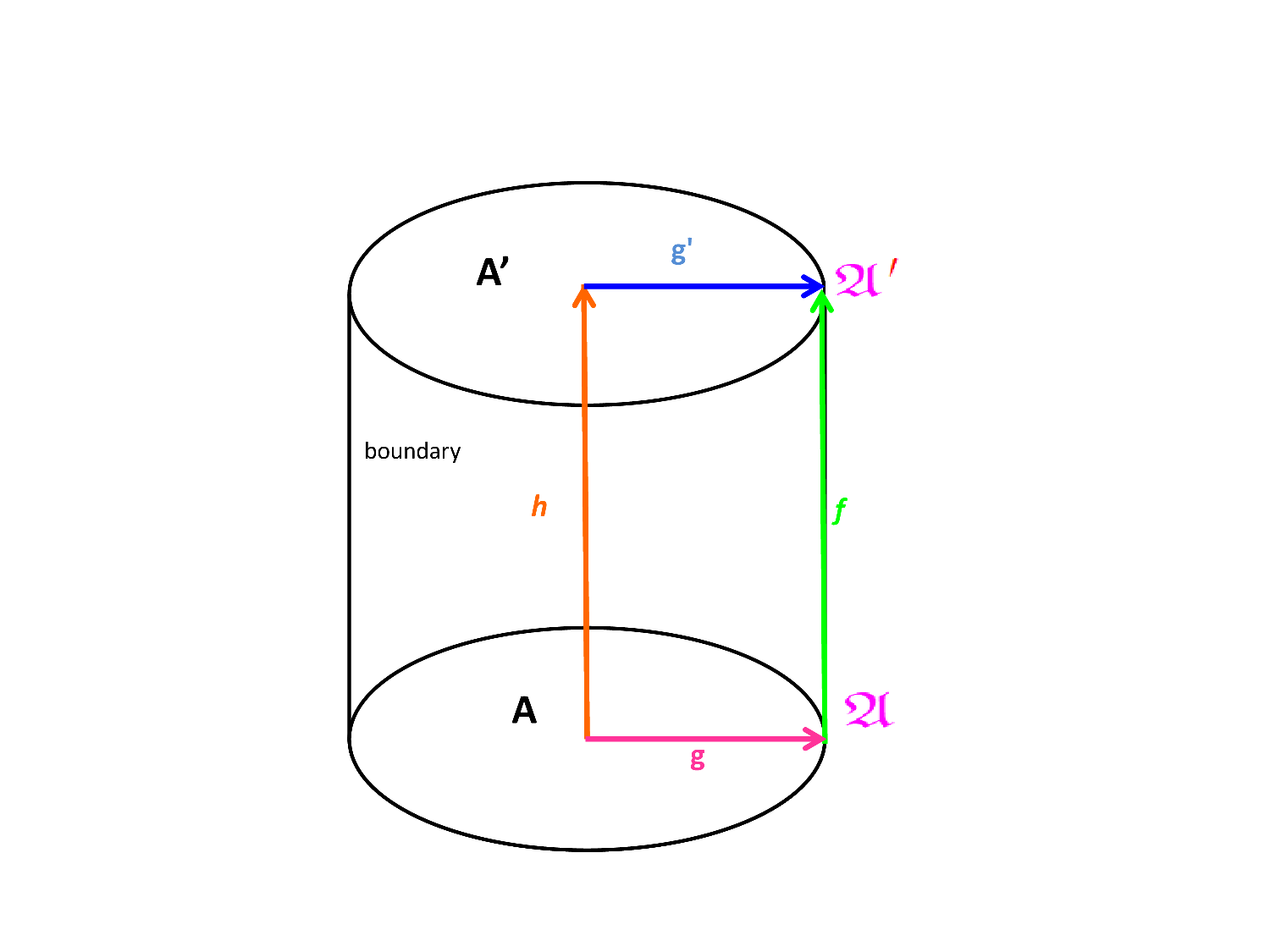}
\end{center}
\par
\vspace{-0.5cm}
\caption{Gauge transformation of boundary gauge fields in AdS$_{3}$ gravity.}
\label{GT}
\end{figure}
namely:%
\begin{equation}
A^{\prime }=\mathfrak{h}^{-1}A\mathfrak{h}+\mathfrak{h}^{-1}d\mathfrak{h,}
\label{hh}
\end{equation}%
where $\mathfrak{h}=h\left( x\right) $ is a bulk gauge group element of $%
SL(2,\mathbb{R})$ that reads as the real exponential $e^{\lambda \left(
x\right) }$ with gauge parameter $\lambda \left( x\right) =K_{a}\lambda
^{a}\left( x\right) $ expanding in the new basis as $K_{+}\lambda
^{+}+K_{0}\lambda ^{0}+K_{-}\lambda ^{-}$. For the component fields $%
A=A_{\mu }dx^{\mu }$, the infinitesimal transformations of $A_{\mu }\left(
x\right) $ read as:
\begin{equation}
\begin{tabular}{lll}
$A_{\mu }^{\prime }$ & $=$ & $\mathfrak{h}^{-1}A_{\mu }\mathfrak{h}+%
\mathfrak{h}^{-1}\partial _{\mu }\mathfrak{h,}$ \\
$A_{\mu }^{\prime }$ & $\simeq $ & $A_{\mu }+\left[ A_{\mu },\lambda \right]
+\partial _{\mu }\lambda +O\left( \lambda ^{2}\right) .$%
\end{tabular}
\label{AA}
\end{equation}%
where in the second line we have used the expansion $\mathfrak{h}\simeq
I+\lambda +O\left( \lambda ^{2}\right) .$ To obtain the transformation of
the gauge potential $\mathfrak{A}$ on the boundary of AdS$_{3}$, we proceed
as follows: $\left( \mathbf{i}\right) $ we consider two boundary gauge
configurations $\mathfrak{A}$ and $\mathfrak{A}^{\prime }$ related to the
bulk gauge fields $A$ and $A^{\prime }$ by two finite gauge transformations $%
\mathfrak{g}=g\left( x\right) $ and $\mathfrak{g}^{\prime }=g^{\prime
}\left( x\right) $\ as in (\ref{H1}). $\left( \mathbf{ii}\right) $ we
replace $A$\ by its expression in terms of the boundary $\mathfrak{A}$
namely $A=\mathfrak{g}^{-1}d\mathfrak{g}+\mathfrak{g}^{-1}\mathfrak{Ag}$, we
end up with the following gauge transformation between the two boundary
gauge field configurations:
\begin{equation}
\mathfrak{A}^{\prime }=\mathfrak{f}^{-1}\mathfrak{Af}+\mathfrak{f}^{-1}d%
\mathfrak{f.}
\end{equation}%
In this transformation, the gauge group element $\mathfrak{f}=\mathfrak{gh}(%
\mathfrak{g}^{\prime })^{-1}$ is given in terms of the gauge parameter $%
\lambda \left( x\right) $ as $\mathfrak{f}=\mathfrak{g}e^{\lambda }(%
\mathfrak{g}^{\prime })^{-1}$ which is an element of $SL\left( 2,\mathbb{R}%
\right) $ that we express like $\mathfrak{f}=e^{\zeta }.$ Infinitesimally,
we have $\delta \mathfrak{f}=\zeta +\mathcal{O}\left( \zeta ^{2}\right) ;$
leading to:%
\begin{equation}
\begin{tabular}{lll}
$\delta \mathfrak{f}$ & $=$ & $\mathfrak{g}\lambda (\mathfrak{g}^{\prime
})^{-1}+\mathcal{O}\left( \lambda ^{2}\right) ,$ \\
& $\equiv $ & $\zeta +\mathcal{O}\left( \zeta ^{2}\right) .$%
\end{tabular}%
\end{equation}%
Here, the infinitesimal gauge parameter $\zeta $ expands like $K_{a}\zeta
^{a}$ and lives on the boundary of AdS$_{3}$ indicating that $\zeta $ is a
function of the boundary coordinates namely $\zeta \left( \xi ^{+},\xi
^{-}\right) $. So, the infinitesimal gauge transformation of the boundary
gauge field is given by:
\begin{equation}
\delta _{\zeta }\mathfrak{A}=d\zeta +\left[ \mathfrak{A},\zeta \right] .
\label{ga}
\end{equation}%
For $\mathfrak{A}_{-}$, we get:%
\begin{equation}
\begin{tabular}{lll}
$\delta _{\zeta }\mathfrak{A}_{-}^{0}$ & $=$ & $\frac{\mathrm{k}}{4\pi }%
\partial _{-}\zeta ^{0}+\mathfrak{A}_{-}^{-}\zeta ^{+}-\mathfrak{A}%
_{-}^{+}\zeta ^{-},$ \\
$\delta _{\zeta }\mathfrak{A}_{-}^{+}$ & $=$ & $-\frac{\mathrm{k}}{2\pi }%
\partial _{-}\zeta ^{+}+2\mathfrak{A}_{-}^{0}\zeta ^{+}+\mathfrak{A}%
_{-}^{+}\zeta ^{0},$ \\
$\delta _{\zeta }\mathfrak{A}_{-}^{-}$ & $=$ & $-\frac{\mathrm{k}}{2\pi }%
\partial _{-}\zeta ^{-}-2\mathfrak{A}_{-}^{0}\zeta ^{-}-\mathfrak{A}%
_{-}^{-}\zeta ^{0}.$%
\end{tabular}
\label{gg2}
\end{equation}%
\textbf{B) Asymptotic algebra}\newline
Under gauge transformations having left moving gauge parameters, the
canonical charge $Q_{{\small \zeta }}^{{\small (-)}}$ of the gauge symmetry
on the boundary of AdS$_{3}$ in terms of $\zeta $ is given by:
\begin{equation}
\begin{tabular}{lll}
$Q_{{\small \zeta }}^{{\small (-)}}$ & $=$ & $\int tr\left( \zeta \mathfrak{A%
}_{-}\right) d\xi ^{-},$ \\
& $=$ & $\int tr\left( \zeta ^{0}\mathfrak{A}_{-}^{0}+\zeta ^{+}\mathfrak{A}%
_{-}^{-}+\zeta ^{-}\mathfrak{A}_{-}^{+}\right) d\xi ^{-}.$%
\end{tabular}
\label{H15}
\end{equation}%
Under this conserved charge, classical fields $\Phi $ transform by the Dirac
bracket $\delta _{\zeta }\Phi =\{\Phi ,Q_{{\small \zeta }}^{{\small (-)}%
}\}_{DB}.$ For the case where $\Phi $ is given by the boundary gauge
potential $\mathfrak{A}_{-}$, we have:%
\begin{equation}
\delta _{\zeta }\mathfrak{A}_{-}=\{\mathfrak{A}_{-},Q_{{\small \zeta }}^{%
{\small (-)}}\}_{DB}.
\end{equation}%
Substituting $Q_{{\small \zeta }}^{{\small (-)}}$ by its value (\ref{H15})
on right side, and using (\ref{gg2}) on the left side leads to the following
affine Kac-Moody (KM) algebra $sl(2,\mathbb{R})_{\mathrm{k}}$:
\begin{equation}
\left\{ \mathfrak{A}_{-}^{a}{\small (\xi }_{1}^{-}{\small )},\mathfrak{A}%
_{-}^{b}{\small (\xi }_{2}^{-}{\small )}\right\} _{DB}=\delta {\small (\xi }%
_{1}^{-}{\small -\xi }_{2}^{-}{\small )}\mathrm{\Upsilon }_{c}^{ab}\mathfrak{%
A}_{-}^{c}{\small (\xi _{2}^{-})+}\frac{\mathrm{k}}{2\pi }{\small \kappa }%
^{ab}\frac{\partial }{\partial \xi _{{\small 1}}^{-}}\delta \left( {\small %
\xi }_{1}^{-}{\small -\xi }_{{\small 2}}^{-}\right) ,  \label{KM}
\end{equation}%
with central extensions given by the anomalous term $\frac{\partial }{%
\partial \xi _{{\small 1}}^{-}}\delta \left( {\small \xi }_{1}^{-}{\small %
-\xi }_{{\small 2}}^{-}\right) .$ \newline
In the end of this derivation we would like to give some useful observations:

\ \

\textbf{C) Comments}\newline
We give two comments; the first regards the full gauge symmetry of the
boundary of AdS$_{3}$; the second concerns the induced conformal invariance.

\begin{itemize}
\item \emph{two copies of Kac-Moody algebras}\newline
Along with the left handed affine KM algebra (\ref{KM}), we also have the
following right handed one:
\begin{equation}
\{\widetilde{\mathfrak{A}}_{+}^{a}{\small (\xi }_{1}^{+}{\small )},%
\widetilde{\mathfrak{A}}_{+}^{b}{\small (\xi }_{2}^{+}{\small )}%
\}_{DB}=\delta {\small (\xi }_{1}^{+}{\small -\xi }_{2}^{+}{\small )}\mathrm{%
\Upsilon }_{c}^{ab}\widetilde{\mathfrak{A}}_{+}^{c}{\small (\xi _{2}^{+})+}%
\frac{\mathrm{\tilde{k}}}{2\pi }{\small \kappa }^{ab}\frac{\partial }{%
\partial \xi _{{\small 2}}^{+}}\delta \left( {\small \xi }_{1}^{+}{\small %
-\xi }_{{\small 2}}^{+}\right) ,  \label{H18}
\end{equation}%
with level $\mathrm{\tilde{k}}=\mathrm{k}{\small \mathrm{.}}$ These KM-
structures define the affine boundary\textrm{\ }of AdS$_{3}$.

\item \emph{Conformal algebra}\newline
The asymptotic symmetry for the GR boundary conditions is given by two
copies of the affine $sl\left( 2\right) _{\mathrm{k}}$-algebras. These KM
symmetries induce a conformal invariance generated by an energy momentum
tensor given by the Sugawara construction. Hence, the algebraic properties
of the topological AdS$_{3}$ gravity are reduced to the 2D conformal
symmetry on the boundary of AdS$_{3}$. In fact, we get the first boundary
conditions that were proposed by Brown and Henneaux \cite{A2} by imposing
the Drinfeld-Sokolov (DS) reduction constraints:
\begin{equation}
\mathfrak{A}_{-}^{+}=-\mathrm{k}/2\pi \text{ \qquad ,\qquad }\mathfrak{A}%
_{-}^{0}=0,
\end{equation}%
then eqs(\ref{H5}) become $\partial _{-}\mathfrak{A}_{+}^{+}=-\mathfrak{A}%
_{+}^{0}$ and $2\mathfrak{A}_{+}^{-}=\partial _{-}\mathfrak{A}_{+}^{0}-\frac{%
2\pi }{\mathrm{k}}\mathfrak{A}_{+}^{+}\mathfrak{A}_{-}^{-}$ as well as:%
\begin{equation}
\partial _{+}\mathfrak{A}_{-}^{-}=\mathfrak{A}_{+}^{+}\partial _{-}\mathfrak{%
A}_{-}^{-}+2\partial _{-}\mathfrak{A}_{+}^{+}\mathfrak{A}_{-}^{-}-\frac{%
\mathrm{k}}{4\pi }\partial _{-}^{3}\mathfrak{A}_{+}^{+},  \label{H22}
\end{equation}%
which is just the integrated version of the Virasoro OPE using Residue
theorem of complex analysis. With similar results on twild sector, the
asymptotic symmetry algebra is no longer the affine $sl\left( 2\right) _{%
\mathrm{k}}$-algebras but it is now given by two copies of the Virasoro
algebra with central charge $c=\tilde{c}=6k=3l_{AdS_{3}}/2G_{N}.$ This
structure defines the conformal boundary\textrm{\ }of AdS$_{3}$.
\end{itemize}

\paragraph{Extended asymptotic symmetry algebra:}

\textrm{In this part, we extend the asymptotic symmetries (\ref{KM}-\ref{H18}%
) by including the current algebra emerging from the unfolded gauge anomaly}%
; particularly from the anomalous boundary term (\ref{bndy}). To engineer
the chiral current $\mathcal{J}_{+}\sim \mathcal{J}_{\bar{z}}$ in the
variation $\delta \mathcal{S}_{tot}^{CS}\left[ A\right] $ (resp. $\widetilde{%
\mathcal{J}}_{-}\sim \widetilde{\mathcal{J}}_{z}$ in the variation $\delta
\widetilde{\mathcal{S}}_{tot}^{CS}$), we start from the 1-form potential $%
A=A_{t}dt+A_{\varphi }d\varphi $ and expresses it like $\mathfrak{A}_{z}d\xi
^{z}+\mathfrak{A}_{\bar{z}}d\xi ^{\bar{z}}$ by applying a Wick rotation $%
t\rightarrow -it$ in (\ref{BC}) where the \textrm{light cone variables} $%
(\xi ^{-},\xi ^{+})$ become $\xi ^{z}=\left( \varphi +it\right) /\sqrt{2}$
and $\xi ^{\bar{z}}=\left( \varphi -it\right) /\sqrt{2}$. For the case of $%
k^{\prime }=-k$ the gauge variation of field action becomes:%
\begin{equation}
\delta \mathcal{S}_{tot}^{CS}=-\frac{k}{2\pi }\int\nolimits_{\partial
\mathcal{M}_{3D}}d^{2}\xi \sqrt{\left\vert \mathrm{h}\right\vert }tr\left[
\delta \mathfrak{A}_{z}\mathfrak{A}_{\bar{z}}\right] .  \label{ano}
\end{equation}%
Substituting $tr\left[ \delta \mathfrak{A}_{z}\mathfrak{A}_{\bar{z}}\right]
=\delta \mathfrak{A}_{z}^{a}\kappa _{ab}\mathfrak{A}_{\bar{z}}^{b},$ the
varied action read as follows:%
\begin{equation}
\begin{tabular}{l}
$\delta \mathcal{S}_{tot}^{CS}=\int\nolimits_{\partial \mathcal{M}%
_{3D}}d^{2}\xi \sqrt{\left\vert \mathrm{h}\right\vert }\mathcal{J}_{\bar{z}%
a}\left( \delta \mathfrak{A}_{z}^{a}\right) $%
\end{tabular}%
,
\end{equation}%
with chiral current:%
\begin{equation}
\mathcal{J}_{\bar{z}a}=-\frac{k}{2\pi }\kappa _{ab}\mathfrak{A}_{\bar{z}%
}^{b}.
\end{equation}%
Applying this analysis to the total AdS$_{3}$ action (\ref{CS}) with CS
level $k^{\prime }=-k$ for the boundary gauge fields $A_{z}$ and CS level $%
\tilde{k}^{\prime }=+\tilde{k}$ for $\tilde{A}_{\bar{z}},$ we obtain:%
\begin{equation}
\mathcal{S}_{CS}^{AdS_{3}}+\mathcal{S}_{CS}^{bnd}=-\frac{k}{2\pi }%
\int\nolimits_{\partial \mathcal{M}_{3D}}\delta \mathfrak{A}_{z}\mathfrak{A}%
_{\bar{z}}+\frac{\tilde{k}}{2\pi }\int\nolimits_{\partial \mathcal{M}%
_{3D}}\delta \widetilde{\mathfrak{A}}_{\bar{z}}\widetilde{\mathfrak{A}}_{z},
\end{equation}%
with chiral currents:
\begin{equation}
\begin{tabular}{lll}
$\mathcal{J}_{\bar{z}a}$ & $=$ & $-\frac{k}{2\pi }\kappa _{ab}\mathfrak{A}_{%
\bar{z}}^{b}$ \\
$\widetilde{\mathcal{J}}_{za}$ & $=$ & $+\frac{\tilde{k}}{2\pi }\kappa _{ab}%
\widetilde{\mathfrak{A}}_{z}^{^{b}}$%
\end{tabular}
\label{anoc}
\end{equation}%
the associated algebra of these currents follows as:%
\begin{eqnarray}
\left\{ \mathcal{J}_{\bar{z}}\left( \xi _{1}^{\bar{z}}\right) _{a},\mathcal{J%
}_{\bar{z}}\left( \xi _{2}^{\bar{z}}\right) _{b}\right\}  &=&\delta {\small %
(\xi }_{1}^{\bar{z}}{\small -\xi }_{2}^{\bar{z}}{\small )}\mathrm{\Upsilon }%
_{c}^{ab}\mathcal{J}_{\bar{z}}^{c}{\small (\xi _{2}^{\bar{z}})}-\frac{k}{%
2\pi }\kappa _{ab}\partial _{\bar{z}}\delta \left( \xi _{1}^{\bar{z}}-\xi
_{2}^{\bar{z}}\right) ,  \label{ex1} \\
\left\{ \mathcal{\tilde{J}}_{z}\left( \xi _{1}^{z}\right) _{a},\mathcal{%
\tilde{J}}_{z}\left( \xi _{1}^{z}\right) _{b}\right\}  &=&\delta {\small %
(\xi }_{1}^{z}{\small -\xi }_{2}^{z}{\small )}\mathrm{\Upsilon }_{c}^{ab}%
\mathcal{\tilde{J}}_{z}^{c}{\small (\xi _{2}^{\bar{z}})}+\frac{\tilde{k}}{%
2\pi }\kappa _{ab}\partial _{z}\delta \left( \xi _{1}^{z}-\xi
_{2}^{z}\right) .  \label{ex2}
\end{eqnarray}%
Taking into account eqs(\ref{KM}-\ref{H18}), the enlarged asymptotic
symmetry for the extended GR boundary conditions is therefore given by:%
\begin{equation}
\begin{tabular}{ccc}
$\left\{ \mathfrak{A}_{-}^{a}{\small (\xi }_{1}^{-}{\small )},\mathfrak{A}%
_{-}^{b}{\small (\xi }_{2}^{-}{\small )}\right\} _{DB}=$ & $\delta {\small %
(\xi }_{1}^{-}{\small -\xi }_{2}^{-}{\small )}\mathrm{\Upsilon }_{c}^{ab}%
\mathfrak{A}_{-}^{c}{\small (\xi _{2}^{-})}$ & ${\small +}\frac{k}{2\pi }%
{\small \kappa }^{ab}\frac{\partial }{\partial \xi _{{\small 1}}^{-}}\delta
\left( {\small \xi }_{1}^{-}{\small -\xi }_{{\small 2}}^{-}\right) ,$ \\
$\left\{ \mathcal{J}_{\bar{z}}\left( \xi _{1}^{\bar{z}}\right) _{a},\mathcal{%
J}_{\bar{z}}\left( \xi _{2}^{\bar{z}}\right) _{b}\right\} =$ &
\multicolumn{1}{l}{$\delta {\small (\xi }_{1}^{\bar{z}}{\small -\xi }_{2}^{%
\bar{z}}{\small )}\mathrm{\Upsilon }_{c}^{ab}\mathcal{J}_{\bar{z}}^{c}%
{\small (\xi _{2}^{\bar{z}})}$} & $-\frac{k}{2\pi }\kappa _{ab}\frac{%
\partial }{\partial \xi _{{\small 1}}^{\bar{z}}}\delta \left( \xi _{1}^{\bar{%
z}}-\xi _{2}^{\bar{z}}\right) ,$%
\end{tabular}%
\end{equation}%
for the left sector. And:%
\begin{equation}
\begin{tabular}{ccc}
$\{\widetilde{\mathfrak{A}}_{+}^{a}{\small (\xi }_{1}^{+}{\small )},%
\widetilde{\mathfrak{A}}_{+}^{b}{\small (\xi }_{2}^{+}{\small )}\}_{DB}=$ & $%
\delta {\small (\xi }_{1}^{+}{\small -\xi }_{2}^{+}{\small )}\mathrm{%
\Upsilon }_{c}^{ab}\widetilde{\mathfrak{A}}_{+}^{c}{\small (\xi _{2}^{+})}$
& ${\small +}\frac{\tilde{k}}{2\pi }{\small \kappa }^{ab}\frac{\partial }{%
\partial \xi _{{\small 2}}^{+}}\delta \left( {\small \xi }_{1}^{+}{\small %
-\xi }_{{\small 2}}^{+}\right) ,$ \\
$\left\{ \mathcal{\tilde{J}}_{z}\left( \xi _{1}^{z}\right) _{a},\mathcal{%
\tilde{J}}_{z}\left( \xi _{1}^{z}\right) _{b}\right\} =$ &
\multicolumn{1}{l}{$\delta {\small (\xi }_{1}^{z}{\small -\xi }_{2}^{z}%
{\small )}\mathrm{\Upsilon }_{c}^{ab}\mathcal{\tilde{J}}_{z}^{c}{\small (\xi
_{2}^{\bar{z}})}$} & $+\frac{\tilde{k}}{2\pi }\kappa _{ab}\frac{\partial }{%
\partial \xi _{{\small 1}}^{z}}\delta \left( \xi _{1}^{z}-\xi
_{2}^{z}\right) ,$%
\end{tabular}%
\end{equation}%
for the right sector. In the end of this section, we give two \textrm{more}
comments regarding: (\textbf{i}) \textrm{the method used in the derivation
of the ASA; we elaborate on our choice of using the variational procedure
instead of the standard method} \textrm{knowing that the two approaches are
equivalent}.\ (\textbf{ii}) the choice of the radial gauge and the potential
loophole.

\begin{description}
\item[$\left( i\right) $] \emph{Comment 1}:\ First notice that eq(\ref{233})
suggests that the charge $Q_{{\small \zeta }}$ can be split as the sum of
a chiral and antichiral charges like $Q_{{\small \zeta }}^{(-)}+Q_{{\small %
\zeta }}^{(+)}$ (or equivalently as $Q_{{\small \zeta }}^{(z)}+Q_{{\small %
\zeta }}^{(\bar{z})}$) with:
\begin{eqnarray}
Q_{{\small \zeta }}^{(z)} &=&-\frac{k}{2\pi }\int tr\left( \zeta \mathfrak{A}%
_{z}\right) d\xi ^{z}=\int \zeta ^{a}\mathcal{J}_{za}d\xi ^{z},  \label{ch}
\\
Q_{{\small \zeta }}^{(\bar{z})} &=&-\frac{k}{2\pi }\int tr\left( \zeta
\mathfrak{A}_{\bar{z}}\right) d\xi ^{\bar{z}}=\int \zeta ^{a}\mathcal{J}_{%
\bar{z}a}d\xi ^{\bar{z}},  \label{hc}
\end{eqnarray}%
where $\mathcal{J}_{za}=-\frac{k}{2\pi }tr\left( K_{a}\mathfrak{A}%
_{z}\right) $ and $\mathcal{J}_{\bar{z}a}=-\frac{k}{2\pi }tr\left( K_{a}%
\mathfrak{A}_{\bar{z}}\right) .$ By using the Killing form; these chiral
currents read respectively as $-\frac{k}{2\pi }\kappa _{ab}\mathfrak{A}%
_{z}^{b}$ and $-\frac{k}{2\pi }\kappa _{ab}\mathfrak{A}_{\bar{z}}^{b}$.
Moreover, following the analysis of subsec \textbf{2.2}, we can derive (\ref%
{ex1}) and (\ref{ex2}) in an identical manner as (\ref{KM}); especially that
both $\mathfrak{A}_{-}$ and $\mathfrak{A}_{+}$ expand analogously in
whatever chosen basis. Still, we wanted to emphasise how the non-invariance
caused by the boundary CS term (\ref{ano}) leads to the computation of the
anomalous chiral currents (\ref{anoc}) and their non-vanishing divergences (%
\ref{S7}) which is why we used the variational procedure instead of the
standard method.

\item[$\left( ii\right) $] \emph{Comment 2:} In \cite{A3}, \textrm{possible
loopholes are due to two main things}: the choice of the radial gauge (\ref%
{CB}-\ref{H1}); and the \textrm{particular} choice of the group element $g$ (%
\ref{H2}). Our work extends the results of \cite{A3} by considering the
non-vanishing variation of the border gauge connection $\mathfrak{A}_{+}$
obtained via a radial gauge map given in (\ref{CB}-\ref{H1}) with the same
group element (\ref{H2}) as in \cite{A3}. Accordingly, our work doesn't
provide any new insights to these potential loopholes but the considerable
arguments provided within \cite{A3} are still applicable for our work.
\end{description}

\section{Higher spins in AdS$_{3}$ gravity}

\label{sec3} In this section, we investigate a generalisation of AdS$_{3}$
gravity by including higher spin gauge fields beyond the 3D graviton. In the
first subsection, we explore the gauge symmetry families allowing such
generalisation. We give a list of higher spin symmetries in AdS$_{3}$
gravity inspired from the Cartan classification of finite dimensional Lie
algebras including, namely $\boldsymbol{A}_{N-1},$ $\boldsymbol{B}_{N},$ $%
\boldsymbol{C}_{N}$ and $\boldsymbol{D}_{N}$, and Tits-Satake graphs of real
forms. In the second subsection, we study higher spins for the\textrm{\ }%
particular case of $SL\left( N,\mathbb{R}\right) $, the real split form of
the $\boldsymbol{A}_{\mathcal{N-}1}$ Lie algebra. For the full case of the
orthogonal $\boldsymbol{B}_{N},$ and $\boldsymbol{D}_{N} $ series, report to
\textrm{\cite{son} }or refer to the appendix for some illustrative examples.

\subsection{Series of higher spin families}

So far, we studied AdS$_{3}$ gravity based on the $SO\left( 2,2\right) $
isometry (\ref{ssp}). Due to the AdS$_{\mathrm{3}}$/CFT$_{\mathrm{2}}$\
correspondence, this 3D gravity theory is often termed as a spin s=2
gravity. By using group homomorphisms, the Anti-de Sitter group $SO\left(
2,2\right) $ can be dealt in several ways drawing the path for various kinds
of generalisations to higher spins AdS$_{3}$\ gravities (s$>$2) as described
below.

\subsubsection{Embedding $SO\left( 2,2\right) $ in higher dim- gauge
symmetries}

\begin{figure}[tbph]
\begin{center}
\includegraphics[width=11cm]{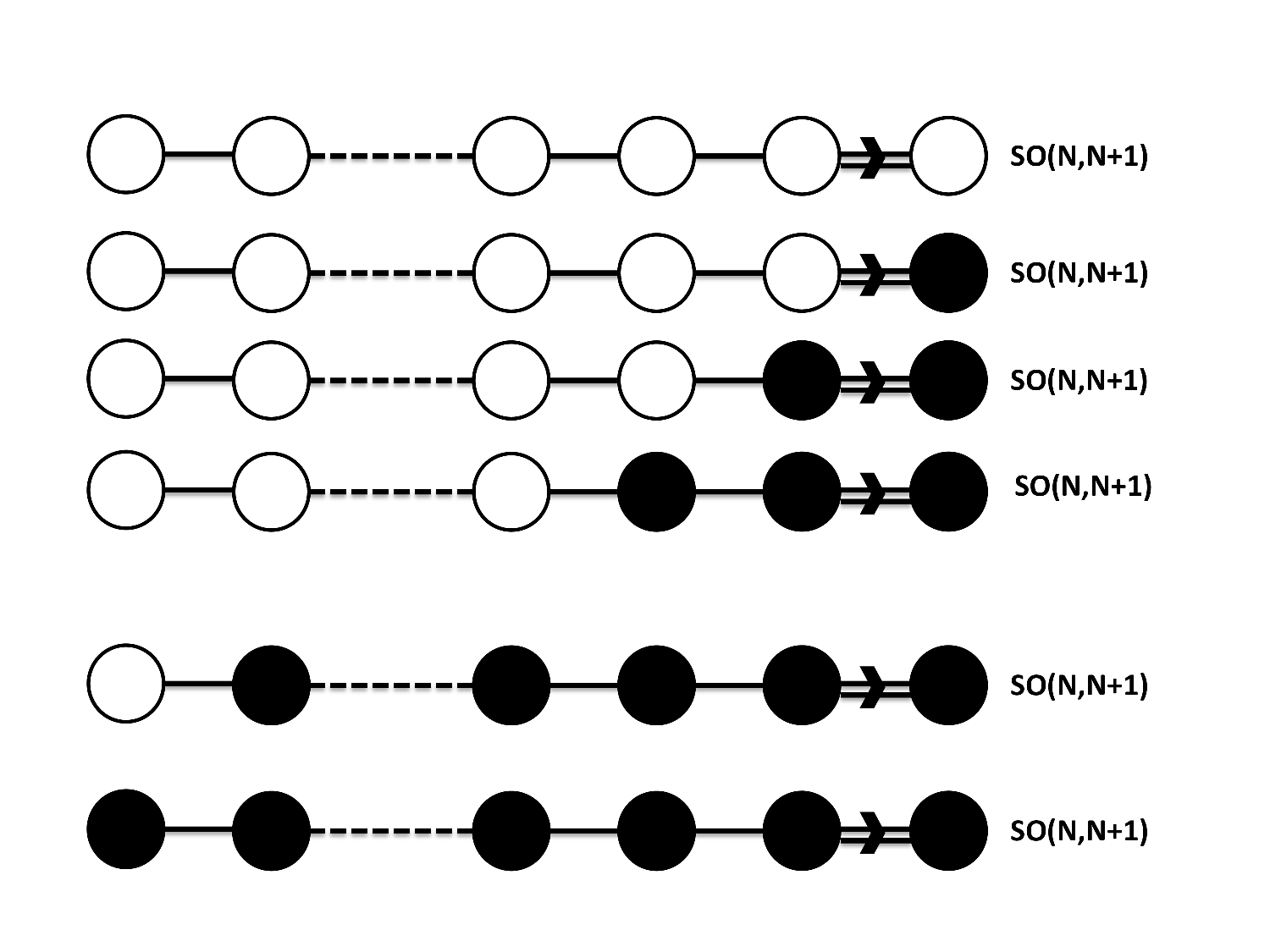}
\end{center}
\par
\vspace{-0.5cm}
\caption{Standard Tits-Satake diagram for the orthogonal family B$_{N}$.}
\label{TSBN}
\end{figure}

We begin by recalling that the two-time signature AdS group $SO\left(
2,2\right) $ can be factorised in terms of two identical copies of the
special linear $SL\left( 2,\mathbb{R}\right) $ groups like:
\begin{equation}
SO\left( 2,2\right) \simeq SL\left( 2,\mathbb{R}\right) _{L}\times SL\left(
2,\mathbb{R}\right) _{R}.  \label{so22}
\end{equation}%
In this realisation, the 3D Lorentzian group $SO\left( 1,2\right) _{X}$ is
thought of in term of the special linear group $SL\left( 2,\mathbb{R}\right)
_{X}$ due to the well known homomorphism $SO\left( 1,2\right) _{X}\sim
SL\left( 2,\mathbb{R}\right) _{X}.$ The 3D Lorentzian group of the usual 3D
spacetime is given by the diagonal subgroup $SO\left( 1,2\right) $ within
the $SO\left( 2,2\right) $ namely:%
\begin{equation}
SO\left( 1,2\right) \simeq \frac{SL\left( 2,\mathbb{R}\right) _{L}\times
SL\left( 2,\mathbb{R}\right) _{R}}{SL\left( 2,\mathbb{R}\right) _{-}}.
\label{eb}
\end{equation}%
Starting from this embedding (\ref{eb}), we can build a family of 3D
gravities with higher spins ($s\geq 2$) by promoting the construction based
on the $SL\left( 2,\mathbb{R}\right) $ gauge to the higher dimensional
groups $SL\left( N,\mathbb{R}\right) $ with $N\geq 2$. In this
generalisation, the 3D Lorentz group appears as the minimal non abelian
subsymmetry as given below:
\begin{equation}
SO\left( 1,2\right) \simeq SL\left( 2,\mathbb{R}\right) \subset SL\left( N,%
\mathbb{R}\right) ,  \label{en}
\end{equation}%
with diagonal $SL\left( N,\mathbb{R}\right) $ as:%
\begin{equation}
SL\left( N,\mathbb{R}\right) \simeq \frac{SL\left( N,\mathbb{R}\right)
_{L}\times SL\left( N,\mathbb{R}\right) _{R}}{SL\left( N,\mathbb{R}\right)
_{-}}.  \label{em}
\end{equation}%
The $sl\left( N,\mathbb{R}\right) $ based construction will be developed
here by using $\left( i\right) $ the principal embedding \textrm{\cite{C7}}
of the Lie algebra $sl\left( 2,\mathbb{R}\right) $ where $sl\left( N,\mathbb{%
R}\right) $ is decomposed into $sl\left( 2,\mathbb{R}\right) $
representations; and $\left( ii\right) $ an interpretation of higher spins
as higher order terms of monomials built with $sl\left( 2,\mathbb{R}\right) $
generators as given in the forthcoming subsection \textbf{3.2}. However,
alternative approaches that implement higher spins are also possible
especially when dealing with manifest $\left( p,q\right) $\ signatures ---
p- time and q- space directions--- like in the orthogonal $SO(p,q)$ and the
symplectic $SP(n,m)$ with:
\begin{equation}
\begin{tabular}{llll}
$SO(p,q)$ & , & $p+q$ & $=2N+1,$ \\
$SO(p,q)$ & , & $p+q$ & $=2N,$ \\
$SP(n,m)$ & , & $n+m$ & $=2N.$%
\end{tabular}%
\end{equation}%
Or when handling exceptional gauge symmetries type the simply laced\textrm{\
}$E_{6,7,8}$. In this regard, an interesting approach to treat higher spins
was described in \cite{son}. It exploits the rich structure of Tits-Satake
root systems of real forms of Lie algebras which are nicely described by the
so-called Tits-Satake graphs \cite{SPIN,son} as depicted by Figure \ref{TSBN}%
.
This approach has been initiated in \cite{son}; and used here through these
Tits-Satake diagrams to study the embedding $SO\left( 2,2\right) $ in higher
dimensional groups G. For that, we will consider gauge symmetries in
relation with the Swampland finiteness conjecture; that is those gauge
groups G with Lie algebras having unbounded ranks like $\boldsymbol{A}_{N-1},
$ $\boldsymbol{B}_{N},$ $\boldsymbol{C}_{N}$ and $\boldsymbol{D}_{N}$. If
instead of the embedding (\ref{so22}), we think about the Anti-de Sitter
group $SO\left( 2,2\right) $ in terms of two Lorentzian groups like:
\begin{equation}
SO\left( 2,2\right) \simeq SO\left( 1,2\right) _{L}\times SO\left(
1,2\right) _{R},
\end{equation}%
one can engineer two standard families of higher spins based on the\textrm{\
}orthogonal Lie algebras $\boldsymbol{B}_{N}$ and $\boldsymbol{D}_{N}.$ By
following \cite{SPIN}, these orthogonal gauge symmetries have $N+1$ standard
real forms and exotic ones that we disregard here. For the $\boldsymbol{B}%
_{N}$ family, we have $\left( i\right) $ the real compact $SO(1+2N)$, $%
\left( ii\right) $ the real split form $SO(N,1+N);$ and $\left( iii\right) $
the $SO(p,q)$ with $p+q=1+2N$ and $p<q.$ In accordance with the principal
embedding of $SL\left( 2,\mathbb{R}\right) $ into $SL\left( N,\mathbb{R}%
\right) ,$ we can generalise $SO\left( 1,2\right) $ by using the real split
form
\begin{equation}
SO(N,1+N),\qquad N\geq 2,  \label{bn}
\end{equation}%
where the 3D Lorentz group appears as the leading member of this orthogonal
series. Regarding the $\boldsymbol{D}_{N}$ family, we also have $\left(
i\right) $ the real compact $SO(2N)$, $\left( ii\right) $ the real split
form $SO(N,N);$ and $\left( iii\right) $ $SO(p,q)$ where $p+q=2N$ with $p<q.$
Here too, the generalisation will be based on the real split form%
\begin{equation}
SO(N,N),\qquad N\geq 2,  \label{dn}
\end{equation}%
having $SO\left( 2,2\right) $ as the leading member and $SL(N,\mathbb{R})$
as a gauge subsymmetry. The full higher spin 3D gravity theories associated
with the orthogonal series (\ref{bn}), (\ref{dn}) and their relationships
with the AdS$_{3}$/CFT$_{2}$\ correspondence as well as the higher spin BTZ
black hole are detailed in a separate work \cite{son}.

\subsubsection{Standard and exceptional classes of higher spins}

Motivated by these higher spin extension ideas in the CS formulation of AdS$%
_{3}$ gravity, and using the Cartan classification of finite dimensional
gauge symmetries as well the Tits-Satake classification of real forms of Lie
algebras, one can conjecture the following family series for the higher spin
3D gravities.

\begin{itemize}
\item \emph{Series of real form of }$SL\left( N,\mathbb{C}\right) $\emph{:}
\newline
This is the special linear family of real forms of the complex Lie algebra $%
\boldsymbol{A}_{N-1}$; it includes the non compact groups $SU\left(
p,q\right) $ with $p+q=N$; the compact $SU\left( N\right) $ and the split
linear real form $SL\left( N,\mathbb{R}\right) .$ For the case of the
standard real forms of complex Lie algebra $sl\left( 4,\mathbb{C}\right) $,
the associated four Tits-Satake diagrams are depicted in Figure\textbf{\ \ref%
{STA3}}.
\begin{figure}[tbph]
\begin{center}
\includegraphics[width=6cm]{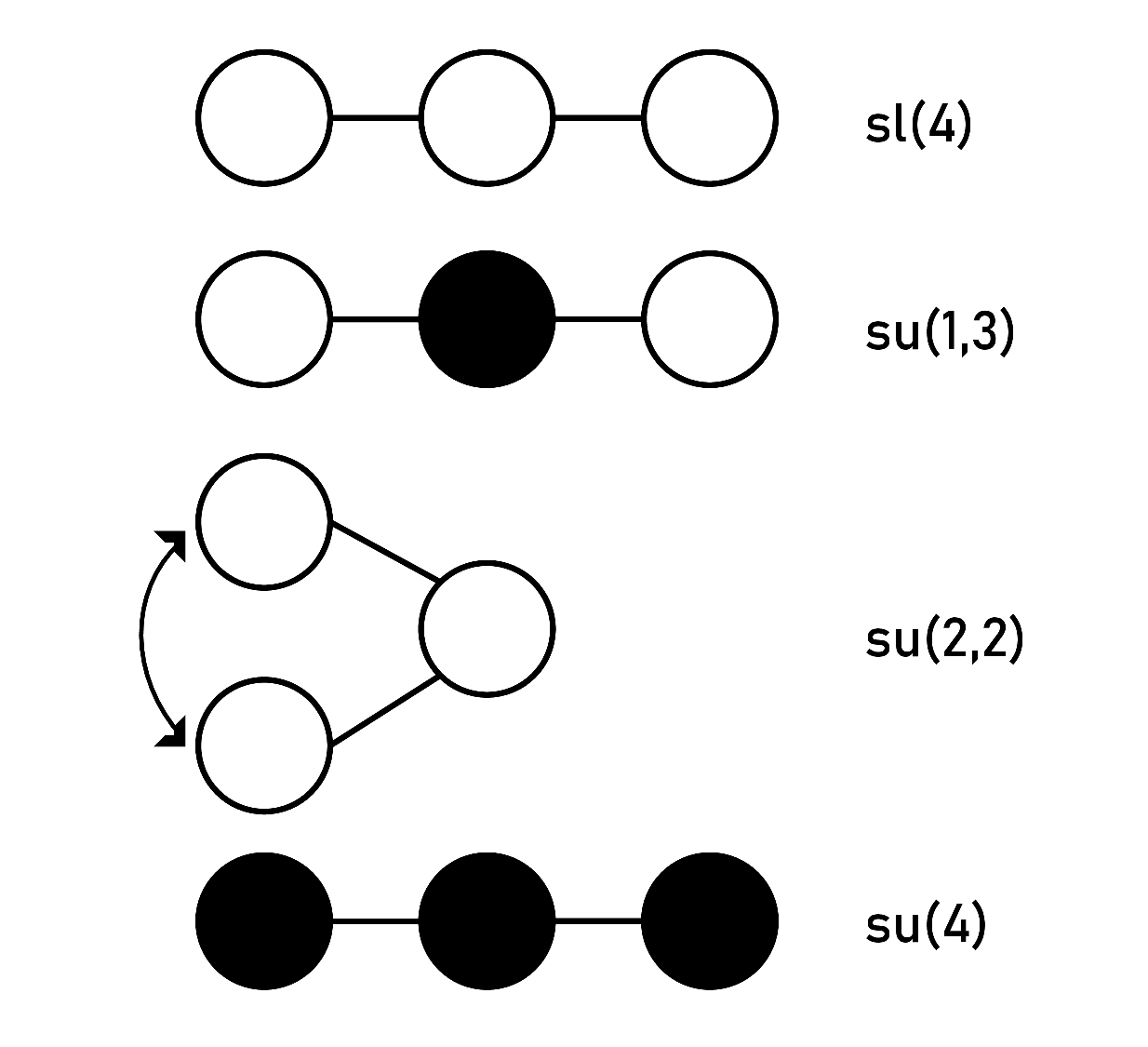}
\end{center}
\par
\vspace{-0.5cm}
\caption{Standard Tits-Satake diagram of $sl(4,\mathbb{R})$.}
\label{STA3}
\end{figure}
In subsection 3.2, we will be particularly interested into the embedding of $%
SO\left( 2,2\right) $ into the sub-family:%
\begin{equation}
SL\left( N,\mathbb{R}\right) _{L}\times SL\left( N,\mathbb{R}\right) _{R}.
\end{equation}%
Using the factorisation:%
\begin{equation}
\begin{tabular}{lll}
$SO\left( 2,2\right) $ & $\simeq $ & $SL\left( 2,\mathbb{R}\right)
_{L}\times SL\left( 2,\mathbb{R}\right) _{R},$ \\
& $\simeq $ & $SL\left( 2,\mathbb{R}\right) _{+}\times SL\left( 2,\mathbb{R}%
\right) _{-},$%
\end{tabular}%
\end{equation}%
the diagonal group $SO\left( 1,2\right) \simeq SL\left( 2,\mathbb{R}\right) $
giving the Lorentzian symmetry of the 3D space is then contained into the
real split form $SL\left( N,\mathbb{R}\right) $ generated by the diagonal
coset:%
\begin{equation}
SL\left( N,\mathbb{R}\right) \simeq \frac{SL\left( N,\mathbb{R}\right)
_{L}\times SL\left( N,\mathbb{R}\right) _{R}}{SL\left( N,\mathbb{R}\right)
_{-}}.
\end{equation}%
For this embedding, the $SL\left( 2,\mathbb{R}\right) $ is interpreted as
just the unpainted (white) node in the Tits-Satake diagram given by Figure
\textbf{\ref{aa}}.
\begin{figure}[tbph]
\begin{center}
\includegraphics[width=8cm]{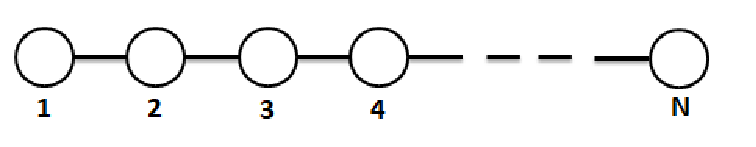}
\end{center}
\par
\vspace{-0.5cm}
\caption{Tits-Satake diagram for the split real forms $sl(N,\mathbb{R})$ of
the Lie algebra $\boldsymbol{A}_{N-1}.$ Each white node describes an sl(2,$%
\mathbb{R}$) subalgebra.}
\label{aa}
\end{figure}

\item \emph{\ Series of real form of }$SO\left( N,\mathbb{C}\right) $
\newline
This is the orthogonal family that contains two particular series namely the
orthogonal sub-family $\boldsymbol{B}_{N}$ and its peer $\boldsymbol{D}_{N}$%
. \newline
The $\boldsymbol{B}_{N}$ class concerns real forms of $SO\left( 1+2N,\mathbb{%
C}\right) $ whose classical ones are given by $SO(p,q)$ with two integer
labels $p$ and $q$ such that $p+q=2N+1.$ An interesting $\boldsymbol{B}_{N}$
series is given by the real split forms (\ref{bn}). \newline
Similarly, the $D_{N}$ class regards real forms of $SO\left( 2N,\mathbb{C}%
\right) $ whose classical ones are also given by $SO(p,q)$ but with two
integer labels $p$ and $q$ verifying $p+q=2N.$ Analogously with (\ref{bn}),
the real split forms of the $D_{N}$ series is given by (\ref{dn}). For this
orthogonal class, the corresponding CS gauge symmetry with higher spins is:
\begin{equation*}
SO(p,q)_{L}\times SO(p,q)_{R}.
\end{equation*}%
For the real split forms $SO(N,1+N)$ and $SO(N,N)$, see the corresponding
Tits-Satake diagram in Figure \textbf{\ref{bd}} below.
\begin{figure}[tbph]
\begin{center}
\includegraphics[width=7cm]{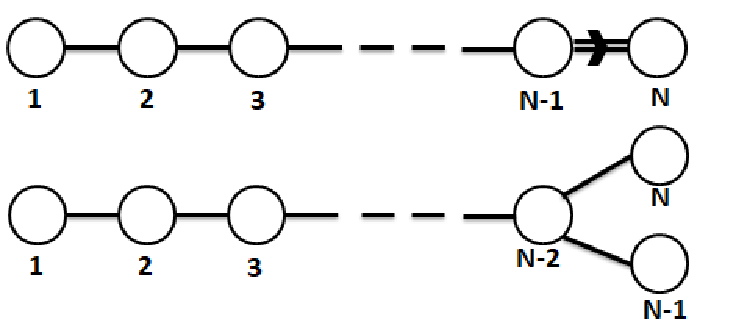}
\end{center}
\par
\vspace{-0.5cm}
\caption{Tits-Satake diagram for: (\textbf{a)-} the split real forms $%
SO(N,N+1)$ of the Lie algebra $\boldsymbol{B}_{N};$ and (\textbf{b)-} the
split real forms $SO(N,N)$ of the Lie algebra $\boldsymbol{D}_{N}.$ Each
node corresponds to $so(1,2).$}
\label{bd}
\end{figure}

\item \emph{\ Series of real form of }$SP\left( N,\mathbb{C}\right) $
\newline
As for the above classes, this symplectic class can be motivated from the
following homomorphism:
\begin{equation}
SO\left( 2,2\right) \simeq Sp\left( 2,\mathbb{R}\right) _{L}\times Sp\left(
2,\mathbb{R}\right) _{R}.
\end{equation}%
Here, the interesting real split form is characterised by one integer label $%
N\geq 1.$ The associated CS gauge symmetry describing higher spin gravity is
given by:%
\begin{equation}
Sp\left( 2N,\mathbb{R}\right) _{L}\times Sp\left( 2N,\mathbb{R}\right) _{R},
\end{equation}%
containing $Sp\left( 2,\mathbb{R}\right) $ as a minimal non abelian
subgroup. The corresponding Tits-Satake diagram is \textbf{\ref{c}}
\begin{figure}[tbph]
\begin{center}
\includegraphics[width=7cm]{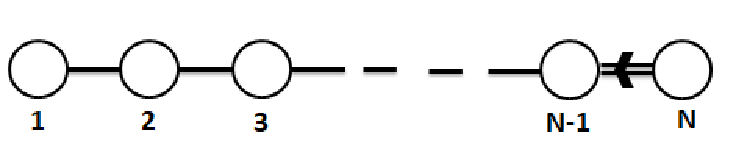}
\end{center}
\par
\vspace{-0.5cm}
\caption{Tits-Satake diagram for the split real forms $sp(2N)$ of the Lie
algebra $\boldsymbol{C}_{N}$}
\label{c}
\end{figure}

\item \emph{Exceptional series}. \newline
This a finite class with exceptional gauge symmetries containing:%
\begin{equation}
\begin{tabular}{lll}
$\mathcal{G}_{2}$ & $:$ & $G_{2}\times \tilde{G}_{2}$ \\
$\mathcal{F}_{4}$ & $:$ & $F_{4}\times \tilde{F}_{4}$%
\end{tabular}%
\qquad ,\qquad
\begin{tabular}{lll}
$\mathcal{E}_{6}$ & $:$ & $E_{6}\times \tilde{E}_{6}$ \\
$\mathcal{E}_{7}$ & $:$ & $E_{7}\times \tilde{E}_{7}$ \\
$\mathcal{E}_{8}$ & $:$ & $E_{8}\times \tilde{E}_{8}$%
\end{tabular}%
,
\end{equation}%
with Tits-Satake diagrams given in Figure \textbf{\ref{e}}. The treatment of
3D gravity Higher spin theory with the simply laced E$_{s}$ family will be
thoroughly described in a separate study.
\begin{figure}[h]
\begin{center}
\includegraphics[width=10cm]{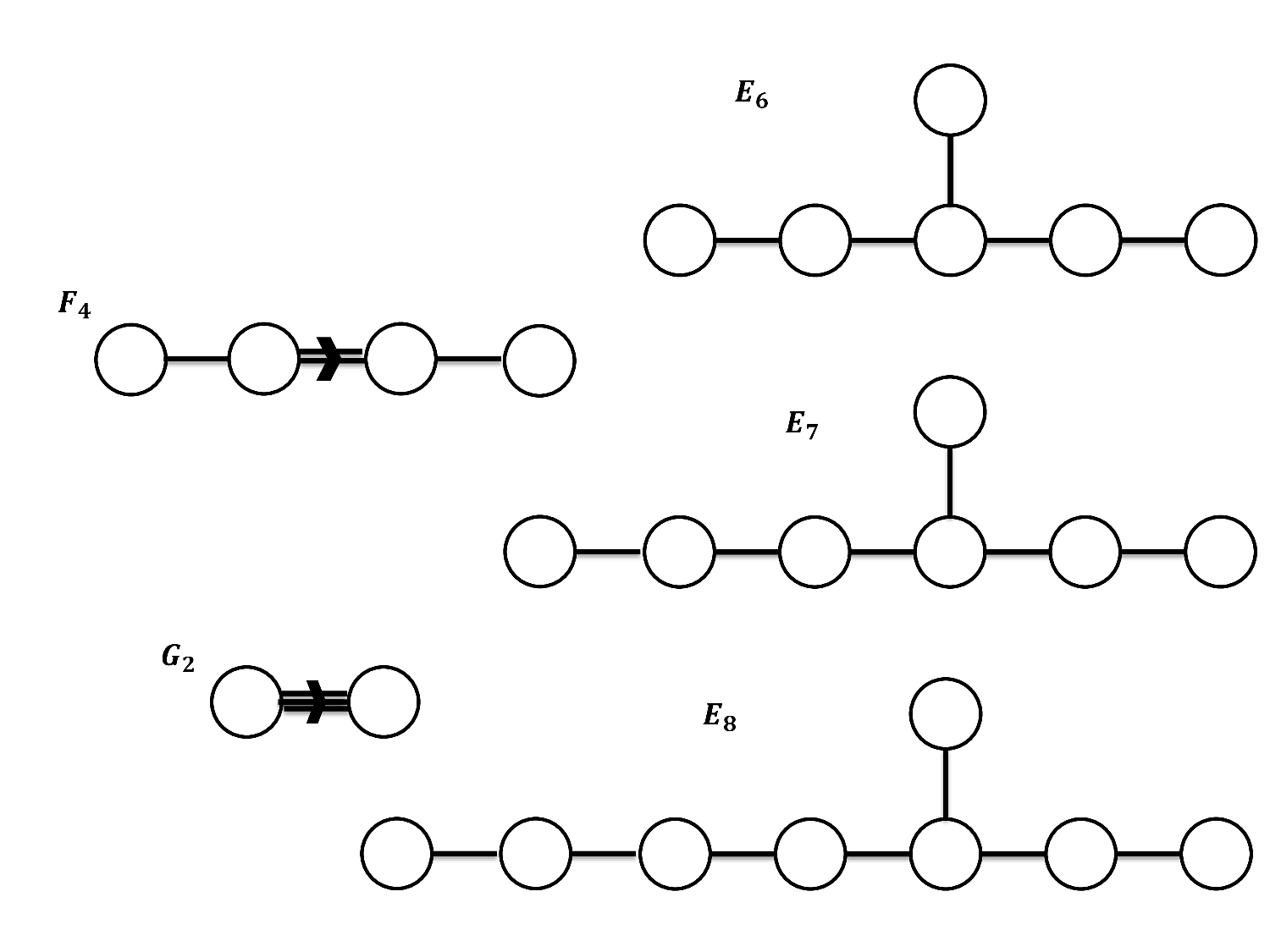}
\end{center}
\par
\vspace{-0.5cm}
\caption{Tits-Satake diagram for the split real forms of exceptional Lie
algebras. All nodes are unpainted.}
\label{e}
\end{figure}
\end{itemize}

\newpage

\subsection{Principal embedding of sl$\left( 2,\mathbb{R}\right) $ in sl$%
\left( N,\mathbb{R}\right) $}

The generalisation of the $sl\left( 2,\mathbb{R}\right) $ theory to the
larger $sl\left( N,\mathbb{R}\right) $ is achieved by decomposing the $%
N^{2}-1$ generators $T_{A}$ of $sl\left( N,\mathbb{R}\right) $ in terms of
monomials $J_{\left( a_{1}...a_{j}\right) }$ of the generators $sl\left( 2,%
\mathbb{R}\right) $ by using the following relation:%
\begin{equation}
N^{2}-1=\sum_{j=1}^{N-1}\left( 2j+1\right) .
\end{equation}%
Concretely, the generators $T_{A}$ are split as follows:
\begin{equation}
T_{A}\qquad \rightarrow \qquad J_{a_{1}}\oplus J_{\left( a_{1}a_{2}\right)
}\oplus ...\oplus J_{\left( a_{1}...a_{N-1}\right) },  \label{dec}
\end{equation}%
where the traceless $J_{\left( a_{1}...a_{n}\right) }\equiv T_{a_{1}...a_{n}}
$ are completely symmetric in the $sl(2,\mathbb{R})$ labels $a_{n}=0,1,2$.
For example, the rank- 2 tensor $J_{\left( ab\right) }$ has the following
quadratic form:%
\begin{equation}
\begin{tabular}{lll}
$J_{\left( ab\right) }$ & $=$ & $\frac{1}{2}\left(
J_{a}J_{b}+J_{b}J_{a}\right) -\frac{\alpha }{2}\eta _{ab}\boldsymbol{J}^{2},$
\\
$\boldsymbol{J}^{2}$ & $=$ & $J_{a}\eta ^{ab}J_{b},$%
\end{tabular}
\label{CJ}
\end{equation}%
with normalisation $\alpha $ determined by the traceless condition $\eta
^{ab}J_{\left( ab\right) }=0$; thus leading to $1-\frac{3\alpha }{2}=0.$
Similarly to the generators $T_{A}$ of $sl\left( N,\mathbb{R}\right) $
obeying the typical commutation relations $\left[ T_{A},T_{B}\right]
=f_{AB}^{C}T_{C};$ the $N-1$ monomials $J_{\left( a_{1}...a_{n}\right) }$
also satisfy closed commutations:%
\begin{equation}
\left[ J_{\left( a_{1}...a_{n}\right) },J_{\left( b_{1}...b_{m}\right) }%
\right] =\sum_{k=1}^{N-1}\mathrm{f}_{\left( a_{1}...a_{n}\right) \left(
b_{1}...b_{m}\right) }^{\left( d_{1}...d_{k}\right) }J_{\left(
d_{1}...d_{k}\right) },
\end{equation}%
with labels $a_{j},b_{j},d_{j}=0,1,2.$ \ By using the short notation $T_{%
\mathbf{a}_{n}}=J_{\left( a_{1}...a_{n}\right) }$ with index vector $\mathbf{%
a}_{n}=\left( a_{1},...,a_{n}\right) $; these commutations relations read as
follows:%
\begin{equation}
\left[ T_{\mathbf{a}_{n}},T_{\mathbf{b}_{m}}\right] =\mathrm{f}_{\mathbf{a}%
_{n}\mathbf{b}_{m}}^{\mathbf{d}_{k}}T_{\mathbf{d}_{k}}.
\end{equation}%
A subset of these commutations is given by:%
\begin{equation}
\begin{tabular}{lll}
$\left[ J_{a},J_{b}\right] $ & $=$ & $\epsilon _{abc}J^{c},$ \\
$\left[ J_{a},T_{a_{1}...a_{n}}\right] $ & $=$ & $\epsilon
_{a(a_{1}}^{d}T_{a_{2}...a_{n})d},$%
\end{tabular}%
\end{equation}%
where in the first relation $\epsilon _{abc}$ is the structure constants of $%
sl\left( 2,\mathbb{R}\right) $. The second relation means that $J_{\left(
a_{1}...a_{n}\right) }$ is a spin $n$ representation of $sl\left( 2,\mathbb{R%
}\right) .$ The other relations are obtained by using eqs like (\ref{CJ})
and the algebra of commutators.\newline
For the example of $sl\left( 3,\mathbb{R}\right) ,$ the eight dimensions
decomposes in two blocks $3+5$ given by the monomials $J_{a}$ and $T_{ab}=%
\frac{1}{2}\left( J_{a}J_{b}+J_{b}J_{a}\right) -\frac{2}{3}\eta _{ab}J^{2}$
obeying:%
\begin{equation}
\begin{tabular}{lll}
$\left[ J_{a},J_{b}\right] $ & $=$ & $\epsilon _{abc}J^{c},$ \\
$\left[ J_{a},T_{bc}\right] $ & $=$ & $\epsilon _{ab}^{d}T_{dc}+\epsilon
_{ac}^{d}T_{bd},$ \\
$\left[ T_{ab},T_{cd}\right] $ & $=$ & $-\left( \eta _{a(c}\epsilon
_{d)bm}+\eta _{b(c}\epsilon _{d)am}\right) J^{m}.$%
\end{tabular}%
\end{equation}%
Notice that the decomposition (\ref{dec}) has an interpretation in terms of
the\textrm{\ }roots of the Lie algebra of $sl(N,\mathbb{R})$ and the
intersecting 2-cycles in complex surfaces with A$_{N-1}$ singularity. For
illustration let us consider the example of $sl\left( 3,\mathbb{R}\right) $
having six roots $\pm \mathbf{\alpha }_{1},$ $\pm \mathbf{\alpha }_{3},$ $%
\pm \mathbf{\alpha }_{3}$ with $\mathbf{\alpha }_{3}=\mathbf{\alpha }_{1}+%
\mathbf{\alpha }_{2}$ and where the $\mathbf{\alpha }_{1}$ and $\mathbf{%
\alpha }_{2}$ are the two simple roots. Using these objects of the Lie
algebra, the eight traceless generators $T_{A}$ of $sl\left( 3,\mathbb{R}%
\right) $ are given by the Cartan charge operators $h_{\mathbf{\alpha }_{1}},
$ $h_{\mathbf{\alpha }_{2}}$ and the six step operators $E_{\pm \mathbf{%
\alpha }_{1}},$ $E_{\pm \mathbf{\alpha }_{2}},$ $E_{\pm \mathbf{\alpha }_{3}}
$. In this Cartan basis, the decomposition (\ref{dec}) may be thought of as
follows:%
\begin{equation}
\begin{tabular}{|l|l|l|}
\hline
$T_{A}$ & Cartan basis & number \\ \hline
$J_{a}$ & $h_{\mathbf{\alpha }_{1}},E_{\pm \mathbf{\alpha }_{1}}$ & 3 \\
\hline
$T_{ab}$ & $h_{\mathbf{\alpha }_{2}},E_{\pm \mathbf{\alpha }_{2}},E_{\pm
\mathbf{\alpha }_{3}}$ & 5 \\ \hline
\end{tabular}
\label{35}
\end{equation}%
where the $J_{a}$ of $sl\left( 2\right) $ has been interpreted in terms of $%
\mathbf{\alpha }_{1}$ while $J_{\left( ab\right) }$ in terms of $\mathbf{%
\alpha }_{2}$ and $\mathbf{\alpha }_{3}$. This choice is not unique, other
splittings are possible; for example by using the generators $h_{\mathbf{%
\alpha }_{3}},E_{\pm \mathbf{\alpha }_{3}}$ associated with the highest root
$\mathbf{\alpha }_{3}$. Another example is given by $sl\left( 4,\mathbb{R}%
\right) ;$ this algebra contains $sl\left( 3,\mathbb{R}\right) ;$ it has 15
dimensions that decompose under $sl\left( 2\right) $ like $3+5+7$
respectively given by the monomials $J_{a},$ $T_{ab}$ and $T_{abc}$ with $%
J_{a},$ $T_{ab}$ as in eq(\ref{35}) while the $T_{abc}$\ given by:%
\begin{equation}
\begin{tabular}{|c|c|c|}
\hline
$sl\left( 4,\mathbb{R}\right) $ & Cartan basis & number \\ \hline
$T_{abc}$ & $h_{\mathbf{\alpha }_{3}},E_{\pm \mathbf{\alpha }_{3}},E_{\pm (%
\mathbf{\alpha }_{2}+\mathbf{\alpha }_{3})},E_{\pm (\mathbf{\alpha }_{1}+%
\mathbf{\alpha }_{2}+\mathbf{\alpha }_{3})}$ & 7 \\ \hline
\end{tabular}%
\end{equation}%
Therefore, within the $sl\left( N,\mathbb{R}\right) $ family, the CS gauge
fields $A_{L}$ and $A_{R}$\ are valued in the $sl\left( N,\mathbb{R}\right) $
Lie algebra; as such they expand in terms of the $sl\left( N,\mathbb{R}%
\right) $ generators $T_{\text{\textsc{A}}}$ like:
\begin{equation}
A_{L}=A_{\mu }^{\text{\textsc{A}}}T_{\text{\textsc{A}}}dx^{\mu }\qquad
,\qquad A_{R}=\tilde{A}_{\mu }^{\text{\textsc{A}}}\tilde{T}_{\text{\textsc{A}%
}}dx^{\mu }.
\end{equation}%
By using the decompositions (\ref{dec}) of the $T_{\text{\textsc{A}}}$'s in
terms of monomials of the $SL\left( 2,\mathbb{R}\right) $ generators namely $%
J_{\left( a_{1}...a_{n}\right) }$, we can present these gauge fields like:%
\begin{equation}
\begin{tabular}{lll}
$\sum\limits_{A=1}^{N^{2}-1}A_{\mu }^{\text{\textsc{A}}}T_{\text{\textsc{A}}}
$ & $=$ & $\sum\limits_{n=1}^{N-1}A_{\mu }^{a_{1}...a_{n}}J_{\left(
a_{1}...a_{n}\right) },$ \\
$A_{\mu }^{a_{1}...a_{n}}$ & $=$ & $\omega _{\mu }^{a_{1}...a_{n}}+\frac{1}{l%
}e_{\mu }^{a_{1}...a_{n}},$%
\end{tabular}
\label{A2}
\end{equation}%
and%
\begin{equation}
\begin{tabular}{lll}
$\sum\limits_{A=1}^{N^{2}-1}\tilde{A}_{\mu }^{\text{\textsc{A}}}T_{\text{%
\textsc{A}}}$ & $=$ & $\sum\limits_{n=1}^{N-1}\tilde{A}_{\mu
}^{a_{1}...a_{n}}J_{\left( a_{1}...a_{n}\right) },$ \\
$\tilde{A}_{\mu }^{a_{1}...a_{n}}$ & $=$ & $\omega _{\mu }^{a_{1}...a_{n}}-%
\frac{1}{l}e_{\mu }^{a_{1}...a_{n}}.$%
\end{tabular}
\label{A3}
\end{equation}%
For convenience, we use below the following short notations:%
\begin{equation}
\begin{tabular}{lll}
$\sum\limits_{A=1}^{N^{2}-1}A_{\mu }^{\text{\textsc{A}}}T_{\text{\textsc{A}}}
$ & $=$ & $\sum\limits_{n=1}^{N-1}A_{\mu }^{\mathbf{a}_{n}}T_{\mathbf{a}%
_{n}},$ \\
$\sum\limits_{A=1}^{N^{2}-1}A_{\mu }^{\text{\textsc{A}}}\tilde{T}_{\text{%
\textsc{A}}}$ & $=$ & $\sum\limits_{n=1}^{N-1}\tilde{A}_{\mu }^{\mathbf{a}%
_{n}}T_{\mathbf{a}_{n}},$%
\end{tabular}%
\end{equation}%
indicating the existence of $N-1$ types of gauge fields $A_{\mu }^{\mathbf{a}%
_{1}},$ $A_{\mu }^{\mathbf{a}_{2}},...,A_{\mu }^{\mathbf{a}_{N-1}}$
associated with the conformal spins $s=2,3,...N$ living on the boundary of
AdS$_{3}$. Moreover, using the Achucarro,Townsend and Witten correspondence
\cite{C1,C2,C7} for 3D gravity, the Chern-Simons description of the higher
spin AdS$_{3}$ gravity action reads as:%
\begin{equation}
\mathcal{S}_{3D}^{\text{hs-grav}}=\mathcal{S}_{CS}^{sl_{N}}\left( A\right) -%
\mathcal{S}_{CS}^{sl_{N}}(\tilde{A}),
\end{equation}%
with gauge potentials given by%
\begin{equation}
A_{\mu }=\sum\limits_{n=1}^{N-1}A_{\mu }^{\mathbf{a}_{n}}T_{\mathbf{a}%
_{n}}\qquad ,\qquad \tilde{A}_{\mu }=\sum\limits_{n=1}^{N-1}\tilde{A}_{\mu
}^{\mathbf{a}_{n}}T_{\mathbf{a}_{n}}.
\end{equation}%
From this decomposition, we learn interesting information as listed below.
\newline
$\left( \mathbf{1}\right) $ The field strengths of these $A_{\mu }$ and $%
\tilde{A}_{\mu }$ are:
\begin{equation}
F_{\mu \nu }=\sum\limits_{n=1}^{N-1}F_{\mu \nu }^{\mathbf{a}_{n}}T_{\mathbf{a%
}_{n}}\qquad ,\qquad \tilde{F}_{\mu \nu }=\sum\limits_{n=1}^{N-1}\tilde{F}%
_{\mu \nu }^{\mathbf{a}_{n}}T_{\mathbf{a}_{n}},
\end{equation}%
with components%
\begin{equation}
\begin{tabular}{lll}
$F_{\mu \nu }^{\mathbf{a}_{n}}$ & $=$ & $\partial _{\mu }A_{\nu }^{\mathbf{a}%
_{n}}-\partial _{\nu }A_{\mu }^{\mathbf{a}_{n}}+\mathrm{f}_{\mathbf{b}_{n}%
\mathbf{b}_{m}}^{\mathbf{a}_{k}}A_{\mu }^{\mathbf{b}_{n}}A_{\nu }^{\mathbf{b}%
_{n}},$ \\
$\tilde{F}_{\mu \nu }^{\mathbf{a}_{n}}$ & $=$ & $\partial _{\mu }\tilde{A}%
_{\nu }^{\mathbf{a}_{n}}-\partial _{\nu }\tilde{A}_{\mu }^{\mathbf{a}_{n}}+%
\mathrm{f}_{\mathbf{b}_{n}\mathbf{b}_{m}}^{\mathbf{a}_{k}}\tilde{A}_{\mu }^{%
\mathbf{b}_{n}}\tilde{A}_{\nu }^{\mathbf{b}_{n}}.$%
\end{tabular}%
\end{equation}%
They also split in terms of $N-1$ types of gauge field strengths $F_{\mu }^{%
\mathbf{a}_{1}},$ $F_{\mu }^{\mathbf{a}_{2}},...,F_{\mu }^{\mathbf{a}_{N-1}}$
and similarly for the twild sector. \newline
$\left( \mathbf{2}\right) $ The CS 3-form given by $\Omega _{3}=AdA+\frac{2}{%
3}A^{3}$ is valued in $sl(N,\mathbb{R})$; and so expands like:
\begin{equation}
\Omega _{3}=\Omega _{3}^{\mathbf{a}_{n}}T_{\mathbf{a}_{n}}\qquad ,\qquad
\Omega _{3}^{\mathbf{a}_{n}}=tr\left( T^{\mathbf{a}_{n}}\Omega _{3}\right) ,
\end{equation}%
with normalisation as $tr\left( T^{\mathbf{a}_{n}}T_{\mathbf{b}_{m}}\right)
=\chi _{n}\delta _{\mathbf{b}_{m}}^{\mathbf{a}_{n}}$. In terms of the gauge
fields, we then have:
\begin{equation}
\Omega _{3}^{\mathbf{a}_{n}}=\kappa _{\mathbf{b}_{m}\mathbf{c}_{p}}^{\mathbf{%
a}_{n}}\left( A^{\mathbf{b}_{m}}dA^{\mathbf{c}_{p}}\right) +\frac{2}{3}%
\kappa _{\mathbf{b}_{m}\mathbf{c}_{p}\mathbf{d}_{q}}^{\mathbf{a}_{n}}\left(
A^{\mathbf{b}_{m}}A^{\mathbf{c}_{p}}A^{\mathbf{d}_{q}}\right) ,
\end{equation}%
with couplings as
\begin{equation}
\kappa _{\mathbf{b}_{m}\mathbf{c}_{p}}^{\mathbf{a}_{n}}=tr\left( T^{\mathbf{a%
}_{n}}T_{\mathbf{b}_{m}}T_{\mathbf{c}_{p}}\right) \qquad ,\qquad \kappa _{%
\mathbf{b}_{m}\mathbf{c}_{p}\mathbf{d}_{q}}^{\mathbf{a}_{n}}=tr\left( T^{%
\mathbf{a}_{n}}T_{\mathbf{b}_{m}}T_{\mathbf{c}_{p}}T_{\mathbf{d}_{q}}\right)
.
\end{equation}%
Similar relations are valid for $\tilde{\Omega}_{3}=\tilde{A}d\tilde{A}+%
\frac{2}{3}\tilde{A}^{3}$. \newline
$\left( \mathbf{3}\right) $ The field equation of motion of the CS field $A$
is given by the vanishing of the gauge curvature; that is $F_{\mu \nu }=0.$
By using the expansion $F_{\mu \nu }=F_{\mu \nu }^{\mathbf{a}_{n}}T_{\mathbf{%
a}_{n}}$; it follows that the curvature in each spin sector $F_{\mu \nu }^{%
\mathbf{a}_{n}}$ must vanish; so we have:
\begin{equation}
F_{\mu \nu }^{\mathbf{a}_{1}}=0,\qquad F_{\mu \nu }^{\mathbf{a}%
_{2}}=0,\qquad ,....,\qquad F_{\mu \nu }^{\mathbf{a}_{N-1}}=0.
\end{equation}%
Similarly for the $SL\left( 2,\mathbb{R}\right) $ case, the $SL\left( N,%
\mathbb{R}\right) $ gauge symmetry maps the gauge potential configuration $%
A_{\mu }$ into an equivalent configuration $A_{\mu }^{\left( h\right) }$
given by:%
\begin{equation}
\partial _{\mu }+A_{\mu }^{\left( h\right) }=h^{-1}\left( \partial _{\mu
}+A_{\mu }\right) h\qquad ,\qquad h\left( x\right) =e^{\lambda \left(
x\right) }\in SL\left( N,\mathbb{R}\right) .
\end{equation}%
Thinking of the boundary of the AdS$_{3}$ as a world sheet surface $\Sigma
_{AdS_{3}}$ parameterised by $\xi ^{\alpha }=\left( t,\varphi \right) $
sitting at $\rho \rightarrow \infty ,$ one can study the value of the $%
A_{\mu }$'s on this surface. We refer to these asymptotic gauge
configurations in terms of 2D gauge potentials $\mathfrak{A}_{\alpha }=%
\mathfrak{A}_{\alpha }\left( \xi \right) $ (boundary gauge field) related to
the bulk gauge field $A_{\mu }$ by an $SL\left( N,\mathbb{R}\right) $ gauge
transformation as follows \cite{C7}:%
\begin{equation}
\begin{tabular}{ccc}
$A_{\alpha }$ & $=$ & $g^{-1}\left( \partial _{\alpha }+\mathfrak{A}_{\alpha
}\right) g,$ \\
$\left. A_{\rho }\right\vert _{\rho \rightarrow \infty }$ & $=$ & constant
matrix C$_{0}.$%
\end{tabular}%
\end{equation}%
Typical gauge transformations realising this feature are given by the
factorisation $g=g_{1}g_{2}$ with $g_{1}=e^{\lambda _{1}}$ and $%
g_{2}=e^{\lambda _{2}}$ where the gauge parameters $\lambda _{1}$ and $%
\lambda _{2}$ are valued in $sl(N,\mathbb{R})$ with the property $\partial
_{\rho }\lambda _{1}=0$ and $\partial _{\rho }\lambda _{2}=C_{0}$. \newline
Focussing on the example of $SL\left( 3,\mathbb{R}\right) ,$ we have the
boundary gauge field:
\begin{equation}
\mathfrak{A}_{\alpha }=\mathfrak{A}_{\alpha }^{\mathbf{b}_{1}}T_{\mathbf{b}%
_{1}}+\mathfrak{A}_{\alpha }^{\mathbf{b}_{2}}T_{\mathbf{b}_{2}},
\end{equation}%
having two spin sectors namely the $\mathfrak{A}_{\alpha }^{\mathbf{b}%
_{1}}T_{\mathbf{b}_{1}}$ for the spin 2 and which corresponds just to the $%
sl\left( 2,\mathbb{R}\right) $ algebra $\mathfrak{A}_{\alpha }^{b}J_{b}$;
and the $\mathfrak{A}_{\alpha }^{\mathbf{b}_{2}}T_{\mathbf{b}_{2}}$
describing the spin 3. The curvature $f_{\alpha \beta }$ associated to the
above boundary fields which is given by $\partial _{\alpha }\mathfrak{A}%
_{\beta }-\partial _{\beta }\mathfrak{A}_{\alpha }+\left[ \mathfrak{A}%
_{\alpha },\mathfrak{A}_{\beta }\right] $ has also two components as follows:%
\begin{equation}
\begin{tabular}{lll}
$f_{t\varphi }^{\mathbf{b}_{1}}$ & $=$ & $\partial _{t}\mathfrak{A}_{\varphi
}^{\mathbf{b}_{1}}-\partial _{\varphi }\mathfrak{A}_{t}^{\mathbf{b}%
_{1}}+\varepsilon _{\mathbf{c}_{1}\mathbf{d}_{1}}^{\mathbf{b}_{1}}\mathfrak{A%
}_{t}^{\mathbf{c}_{1}}\mathfrak{A}_{\varphi }^{\mathbf{d}_{1}},$ \\
$f_{t\varphi }^{\mathbf{b}_{2}}$ & $=$ & $\partial _{t}\mathfrak{A}_{\varphi
}^{\mathbf{b}_{2}}-\partial _{\varphi }\mathfrak{A}_{t}^{\mathbf{b}%
_{2}}+\varepsilon _{\mathbf{c}_{2}\mathbf{d}_{2}}^{\mathbf{b}_{2}}\mathfrak{A%
}_{t}^{\mathbf{c}_{2}}\mathfrak{A}_{\varphi }^{\mathbf{d}_{2}}.$%
\end{tabular}%
\end{equation}%
Depending on the choice of the boundary conditions, we distinguish two types
of asymptotic symmetries.

\begin{description}
\item[$\left( \mathbf{i}\right) $] the Kac-Moody symmetry $SL\left(
3,R\right) _{k_{L}}\times SL\left( 3,R\right) _{k_{R}}$ ; and the current
algebra $\overline{SL\left( 3,R\right) }_{k_{L}}\times \overline{SL\left(
3,R\right) }_{k_{R}}$ with CS levels $\left( k_{L},k_{R}\right) $ for the
affine boundary.

\item[$\left( \mathbf{ii}\right) $] the $W_{3}\times \bar{W}_{3}$ invariance
with central charges $\left( c,\bar{c}\right) =\left( 6k,6\bar{k}\right) $
for the conformal boundary.
\end{description}

For generic $N,$ the 2D field theory pertaining on the boundary of AdS$_{3}$
has rich symmetries; in particular: $\left( i\right) $ an affine KM symmetry
$SL\left( N,R\right) _{k_{L}}\times SL\left( N,R\right) _{k_{R}}$ and the
current algebra $\overline{SL\left( N,R\right) }_{k_{L}}\times \overline{%
SL\left( N,R\right) }_{k_{R}}$ behind the Lax- integrability of exactly
solvable 2D systems. $\left( ii\right) $ a non linear W-invariance $W\left(
sl_{N}\right) \times \bar{W}\left( sl_{N}\right) $ that can be imagined in
terms of higher order Casimirs of the KM Lie algebra with central charges $%
\left( c,\bar{c}\right) =\left( 6k,6\bar{k}\right) $.

\section{Swampland constraints on HS gravity Landscape}

\label{sec4} Now that \textrm{we have established }the Chern-Simons
formulation of the 3D anti- de Sitter space gravity and its dual asymptotic
conformal field theory, \textrm{we use} the previous classification of
higher spin families \textrm{to define} an AdS$_{3}$ Landscape (3D
Landscape). Afterwards, \textrm{we study }the anomalies at the 2D boundary
\cite{C, S1, S2} \textrm{to verify} the Swampland BNMM conjecture given by
the upper bound on the rank of the gauge symmetry groups associated to the
higher spin gravity theories. \textrm{All in all to endorse the finiteness
of the 3D Landscape}. In subsection \textbf{4.1}, we \textrm{introduce} the%
\textrm{\ }3D\textrm{\ }Landscape. In subsection \textbf{4.2}, we derive the
anomaly polynomial of the boundary CFT$_{2}$ by evaluating the currents that
arise from the gauge anomaly; then we determine the non conservation of the
stress tensor that ensues from the gravitational anomaly. After, we discuss
the anomaly inflow mechanism for the cancellation of the anomaly polynomial.
In subsection \textbf{4.3}, we verify the Swampland conjecture in AdS$_{3}$\
gravity by computing the upper bound on the rank of the higher spin gauge
symmetries of the 3D Landscape. In subsection \textbf{4.4}, we intersect the
finiteness and the AdS distance conjectures to sharpen the bound and expand
the definition of the AdS$_{3}$ Landscape. In subsection \textbf{4.5}, we
compare the derived Swampland bound with the gravitational exclusion
principle \cite{G2} and excerpt the limitations on the CS level \textrm{k}.

\subsection{Landscape of AdS$_{3}$ gravity}

As defined in \cite{A12222, A1222}, the Swampland theories are a set of
effective field theories\textrm{\ }that cannot be completed consistently
into a UV quantum gravity theory. This consistency is determined by a set of
criteria known as Swampland conjectures. Oppositely, Landscape theories form
a set of effective field theories that verify the Swampland conjectures and
thus are consistent and can be completed into UV quantum gravitational
theories. In our current work, we focus on one particular criterion, the
finiteness of the Landscape constraint, to distinguish between the 3D
Swampland and 3D Landscape theories. Henceforward, we designate by 3D
Landscape theories the set of higher spin AdS$_{3}$ gravity theories from
the classification in subsection \textbf{3.1} of higher spin families that
verify the finiteness conjecture. Therefore, the provisional AdS$_{3}$
Landscape theories with HS given by real split forms are characterised by
the following gauge symmetries
\begin{equation}
\begin{tabular}{c}
AdS$_{3}$ Landscape for real split forms \\ \hline\hline
$SL\left( N,\mathbb{R}\right) _{L}\times SL\left( N,\mathbb{R}\right) _{R}$
\\ \hline
$SO(N,1+N)_{L}\times SO(N,1+N)_{R}$ \\ \hline
$SO(N,N)_{L}\times SO(N,N)_{R}$ \\ \hline
$Sp\left( 2N,\mathbb{R}\right) _{L}\times Sp\left( 2N,\mathbb{R}\right) _{R}$
\\ \hline
$G_{2(2)}\times G_{2(2)}$ \\ \hline
$F_{4(4)}\times F_{4(4)}$ \\ \hline
$E_{6(6)}\times E_{6(6)}$ \\ \hline
$E_{7(7)}\times E_{7(7)}$ \\ \hline
$E_{8(8)}\times E_{8(8)}$ \\ \hline\hline
\end{tabular}%
\end{equation}%
\begin{equation*}
\end{equation*}%
Notice that the finiteness of the AdS$_{3}$ Landscape depend on the
finiteness of the higher spin gauge families based on the real split forms $%
sl(N,\mathbb{R})$, $so(N,1+N)$, $so(N,N)$ and $sp(2N,\mathbb{R})$. For
finite dimensional gauge groups with bounded N, the associated 3D Landscape
must be also finite. Hence, the Swampland finiteness conjecture given by the
upper bound on the rank of possible HS-AdS$_{3}$ gauge groups implies the
finiteness of the 3D Landscape.

\subsection{Anomalies on the 2D boundary}

Here, we study the anomalies originating from boundary terms of HS-AdS$_{3}$
gravity. The anomaly polynomial $\mathcal{I}_{4}\left( F,R\right) $ in terms
of the gauge $F$ and the gravitational $R$ curvatures living on the 2D
boundary of AdS$_{3}$ splits as:
\begin{equation}
\mathcal{I}_{4}[F,R]=\mathcal{I}_{4}^{gauge}[F]+\mathcal{I}_{4}^{grav}[R],
\label{poly}
\end{equation}%
where $\mathcal{I}_{4}^{gauge}[F]$ is the gauge anomaly and $\mathcal{I}%
_{4}^{grav}[R]$ is its gravitational homologue; both originating from the
defective variational principle of $\mathcal{S}_{AdS_{3}}^{tot}=\mathcal{S}_{%
{\small CS}}^{gauge}+\mathcal{S}_{{\small CS}}^{grav}.$ The additional
gravitational Chern-Simons term $\mathcal{S}_{{\small CS}}^{grav}$ ensues
from requiring different left and right central charges ($c_{L}\neq c_{R}$).
The 4-form $\mathcal{I}_{4}[F,R]$ is given by $\mathcal{I}_{4}=d\mathcal{I}%
_{3}$ where $\mathcal{I}_{3}$\ decomposes like:%
\begin{equation}
\begin{tabular}{lll}
$\mathcal{I}_{3}$ & $=$ & $\mathcal{I}_{3}^{gauge}[A,F]+\mathcal{I}%
_{3}^{grav}[\Gamma ,R],$ \\
$\mathcal{I}_{4}^{gauge}$ & $=$ & $d\mathcal{I}_{3}^{gauge}[A,F],$ \\
$\mathcal{I}_{4}^{grav}$ & $=$ & $d\mathcal{I}_{3}^{grav}[\Gamma ,R],$%
\end{tabular}
\label{SI1}
\end{equation}%
and it is related to the total CS field action as:
\begin{equation}
\mathcal{S}_{AdS_{3}}^{tot}=\int_{\mathcal{M}_{3D}}\mathcal{I}_{3}.
\end{equation}%
In these relations, the $A$ is the gauge potential $dx^{\mu }A_{\mu }^{\text{%
\textsc{a}}}T_{\text{\textsc{a}}}$ and the $\Gamma _{\alpha }^{\beta }$ is
the Levi-Civita 1-form $dx^{\mu }\Gamma _{\mu \alpha }^{\beta }.\ $The
variation of the total field action is anomalous; it varies as:
\begin{equation}
\delta \mathcal{S}_{AdS_{3}}^{tot}=\int_{\partial \mathcal{M}_{3D}}\mathcal{I%
}_{2}\qquad ,\qquad \delta \mathcal{I}_{3}=d\mathcal{I}_{2},  \label{SI}
\end{equation}%
with 2-form $\mathcal{I}_{2}$ splitting as $\mathcal{I}_{2}^{gauge}+\mathcal{%
I}_{2}^{grav}$ respectively related to the non vanishing divergences of the
gauge current $\nabla ^{\mu }\mathcal{J}_{\mu }^{a}$ and the energy momentum
tensor $\nabla _{\mu }T^{\mu \nu }.$

\subsubsection{Gauge and gravitational anomalies}

We first study the gauge anomaly, then we consider the gravitational anomaly
induced by the dis-symmetry of the left and right central charges $c_{L}$ and
$c_{R}.$

\paragraph{Gauge anomaly:}

Following the elaboration in subsection \textbf{2.3, }the gauge anomaly
induces chiral currents at the boundary of the form:%
\begin{equation}
\begin{tabular}{lll}
$\mathcal{J}_{\bar{z}a}$ & $=$ & $\frac{k}{2\pi }\kappa _{ab}\mathfrak{A}_{%
\bar{z}}^{b},$ \\
$\widetilde{\mathcal{J}}_{za}$ & $=$ & $-\frac{\tilde{k}}{2\pi }\kappa _{ab}%
\widetilde{\mathfrak{A}}_{z}^{^{b}},$%
\end{tabular}%
\end{equation}%
with divergences given by
\begin{equation}
\begin{tabular}{lllllll}
$\nabla _{\bar{z}}\widetilde{\mathcal{J}}_{za}$ & $=$ & $-\frac{\tilde{k}}{%
2\pi }\kappa _{ab}(\nabla _{\bar{z}}\widetilde{\mathfrak{A}}_{z}^{^{b}})$ & $%
\qquad ,\qquad $ & $\nabla _{z}\mathcal{J}_{\bar{z}a}$ & $=$ & $\frac{k}{%
2\pi }\kappa _{ab}\left( \nabla _{z}\mathfrak{A}_{\bar{z}}^{b}\right) .$%
\end{tabular}
\label{S7}
\end{equation}%
Generally, the 2D chiral anomaly \cite{S1,S2} conveys the violation of the
conservation of chiral currents as follows:%
\begin{equation}
\nabla ^{\mu }\mathcal{J}_{\mu a}=\frac{q_{ab}}{8\pi }F_{\mu \nu
}^{b}\varepsilon ^{\mu \nu },  \label{S8}
\end{equation}%
where $q_{ab}$ are 't Hooft anomaly coefficients such that\emph{\ }$%
q_{ab}=-\left( q_{ab}\right) _{L}$ for $\left( a,b\right) $ both left moving
and $q_{ab}=+\left( q_{ab}\right) _{R}$ if $\left( a,b\right) $ are both
right moving. Comparing (\ref{S8}) with (\ref{S7}), we conclude that the CS
level $k$ is equivalent to the matrix of 't Hooft anomaly coefficients. More
precisely, we have:%
\begin{equation}
q_{ab}=\pm 4k\kappa _{ab}.
\end{equation}%
With the scheme (\ref{SI1}-\ref{SI}), eq(\ref{S7}) constitutes the gauge
sector $\mathcal{I}_{4}^{gauge}[F]$ of the total anomaly polynomial $%
\mathcal{I}_{4}[F,R].$

\paragraph{Gravitational anomalies:}

Here, we will adopt a different approach to derive the gravitational anomaly
$\mathcal{I}_{4}^{grav},$ compared to the computation of the gauge anomaly.
This approach is known as the anomaly inflow mechanism \cite{inflow}; it
describes the cancellation of anomalies by adding the contribution from the
border CFT such that:
\begin{equation}
\mathcal{I}_{4}=-\mathcal{I}_{4}^{\text{inflow}},
\end{equation}%
with $\mathcal{I}_{4}$ given by (\ref{poly}). Particularly, we focus on:%
\begin{equation}
\mathcal{I}_{4}^{grav}=-\mathcal{I}_{4}^{grav,\text{ inflow}}.
\end{equation}%
In order to derive $\mathcal{I}_{4}^{grav}$ directly from the Chern-Simons
action, \textrm{one need to infer a description of this action under a
diffeomorphisms invariance; so we will use results from the dual CFT, namely
}$I_{4}^{grav,\text{ inflow}}$\textrm{\ to deduce the form of the
gravitational anomaly, it is given by} \cite{GA}%
\begin{equation}
\partial _{\mu }T^{\mu \nu }=\frac{c_{L}-c_{R}}{192\pi }\epsilon ^{\nu \mu
}\partial _{\mu }R.  \label{gravinflow}
\end{equation}%
From the perspective of AdS$_{3}$, the gravitational anomaly $\mathcal{I}%
_{4}^{grav}$ must be therefore engendered by an additional CS term that
account for the difference $c_{R}\neq c_{L}$ \cite{c}:%
\begin{equation}
\mathcal{S}_{CS}^{grav}\left( \Gamma \right) =-\frac{c_{L}-c_{R}}{96\pi }%
\int_{\mathcal{M}_{3D}}Tr\left( \Gamma d\Gamma +\frac{2}{3}\Gamma
^{3}\right) ,  \label{gama}
\end{equation}%
the variation of the stress tensor follows from the breakdown of
diffeomorphisms invariance $\delta \Gamma $ and is given as a sign flip of (%
\ref{gravinflow}), we consequently get:%
\begin{equation}
\partial _{\mu }T^{\mu \nu }=\frac{c_{R}-c_{L}}{96\pi }g^{\nu \alpha
}\epsilon ^{\mu \rho }\partial _{\beta }\partial _{\mu }\Gamma _{\alpha \rho
}^{\beta },  \label{S10}
\end{equation}%
with $R_{\alpha \mu \rho }^{\beta }=2\partial _{\mu }\Gamma _{\left[ \alpha
\rho \right] }^{\beta }$. The shift in the central charges should be of the
same magnitude but with opposite signs \cite{c}, therefore:%
\begin{equation}
\begin{tabular}{lll}
$c_{L}-c_{R}$ & $=$ & $96\pi \beta ,$ \\
$c_{L}$ & $=$ & $\frac{3l}{2G}+48\pi \beta ,$ \\
$c_{R}$ & $=$ & $\frac{3l}{2G}-48\pi \beta .$%
\end{tabular}
\label{CLR}
\end{equation}%
Using (\ref{SI1}-\ref{SI}), the total anomaly polynomial for both gauge and
gravitational non invariance is therefore a 4-form that we can write as the
following \cite{inflow}:%
\begin{equation}
\mathcal{I}_{4}^{\text{inflow}}=\frac{1}{2}\sum_{a,b}q_{ab}\boldsymbol{c}%
_{1}\left( F^{a}\right) \wedge \boldsymbol{c}_{1}\left( F^{b}\right) -\frac{%
c_{R}-c_{L}}{24}p_{1}\left( R\right) ,  \label{inflow}
\end{equation}%
where $\boldsymbol{c}_{1}\left( F\right) $ is the first Chern class and $%
p_{1}\left( R\right) $ is the first Pontryagin class. It can be also
expressed like:%
\begin{equation}
\mathcal{I}_{4}^{\text{inflow}}=-\sum_{a,b}\frac{q_{ab}}{8\pi ^{2}}%
F^{a}F^{b}+\frac{c_{R}-c_{L}}{192\pi ^{2}}TrR^{2}.  \label{4inf}
\end{equation}%
Hence,%
\begin{equation}
\mathcal{I}_{4}=\sum_{a,b}\frac{k}{2\pi ^{2}}F^{a}\kappa _{ab}F^{b}-\frac{%
c_{R}-c_{L}}{192\pi ^{2}}TrR^{2}.
\end{equation}%
In order for the theory to be consistent, i.e free of anomalies, $\mathcal{I}%
_{4}$ must be cancelled by $\mathcal{I}_{4}^{\text{inflow}}$. As seen above,
$\mathcal{I}_{4}$ arises from the gauge and gravitational Chern-Simons terms
that we have added along the discussion; either to derive the boundary gauge
currents or to differentiate between the left and right central charges ($%
c_{R}\neq c_{L}$). As for $\mathcal{I}_{4}^{\text{inflow}}$, one must ask
about the origin of these boundary degrees of freedom that generate the
anomalies within to justify its existence and therefore the cancellation $%
\mathcal{I}_{4}+\mathcal{I}_{4}^{\text{inflow}}=0.$ As we will see below,
the $\mathcal{I}_{4}^{\text{inflow}}$ is due to strings.

\subsubsection{Strings at the AdS$_{3}$ boundary}

Usually, the 4-form $\mathcal{I}_{4}$ anomaly polynomial cancels from the
worldsheet degree of freedom of a string as in \cite{C5, C6}. Similarly, one
might ponder about the existence of such string at the boundary of AdS$_{3}$
to explain the emergence of $\mathcal{I}_{4}^{\text{inflow}}$ (\ref{4inf})$.$
Indeed, It has been stated in various occasions, see for instance \cite%
{wzw1,wzw2,wzw4}, that the Chern-Simons theory $\mathcal{S}_{{\small CS}}^{%
{\small G}}\left( \mathbf{A}\right) $ with symmetry \textbf{G} on a manifold
$\boldsymbol{M}$ with boundary ($\partial \boldsymbol{M}\neq \emptyset $) is
d\textrm{ual to a Wess-Zumino-Witten theory with action }$\mathcal{S}_{%
{\small WZW}}^{{\small G}}\left[ \mathbf{g}\right] $ whose expression will
be given later. This property is often known as the CS/WZW duality. For the
matter at hand, the AdS$_{3}$ gravity given by the difference of two CS
actions $\mathcal{S}_{{\small CS}}^{{\small G}_{L}}-\mathcal{S}_{{\small CS}%
}^{{\small G}_{R}}$ as in (\ref{CS}), can be expressed as a difference of
two chiral WZW\ actions $\mathcal{S}_{{\small WZW}}^{{\small G}_{L}}-%
\mathcal{S}_{{\small WZW}}^{{\small G}_{R}}$; defining therefore a non
chiral WZW action $\mathcal{S}_{wzw}^{NC}\left[ g\right] $ with $%
g=g_{L}^{-1}g_{R}$. Taking advantage of this duality, and the fact that we
can define a string as a 2D CFT \cite{sky}, we introduce a string in our AdS$%
_{3\text{ }}$boundary via the non-chiral WZW action $\mathcal{S}_{wzw}^{NC}%
\left[ g\right] $ having two block terms: $\left( \mathbf{i}\right) $ a bulk
term $\mathcal{S}_{bulk}\left[ g\right] $ with gauge group element $g\left(
x\right) $ living in AdS$_{3}$ with coordinates $\left( x\right) ;$ and $%
\left( \mathbf{ii}\right) $ a boundary surface term $\mathcal{S}_{edge}\left[
g\right] $ with $g\left( \xi \right) $ living at the boundary $\partial
(AdS_{3})$ with coordinates $\left( \xi \right) $. These two terms read
explicitly as follows:%
\begin{equation}
\mathcal{S}_{wzw}^{NC}\left[ g\right] =\mathcal{S}_{bulk}\left[ g\right] +%
\mathcal{S}_{edge}\left[ g\right] ,  \label{NCWZW}
\end{equation}%
with bulk term as%
\begin{equation}
\mathcal{S}_{bulk}\left[ g\right] =\frac{k}{12\pi }\int_{\boldsymbol{M}%
}Tr\left( g^{-1}dg\right) ^{3},
\end{equation}%
where we have set $\boldsymbol{M}=AdS_{3};$ and boundary term having the
typical form:%
\begin{equation}
\mathcal{S}_{edge}\left[ g\right] =\frac{k}{4\pi }\int_{\partial \boldsymbol{%
M}}Tr\left[ \left( g^{-1}\nabla _{z}g\right) \left( g^{-1}\nabla _{\bar{z}%
}g\right) \right] .
\end{equation}%
To exhibit the string term in the boundary field action, let us focus on the
diagonal parts $g^{-1}\partial _{z}g$ and $g^{-1}\partial _{\bar{z}}g$
involved in the gauge covariant derivative block terms making $\mathcal{S}%
_{edge}\left[ g\right] .$ For that, we parameterise the invertible gauge
group elements like $g=\exp \left( X\right) $ with gauge field parameter
matrix $X$ valued in the Lie algebra of the gauge symmetry $G$; this matrix $%
X$ is given by the expansion $\sum T_{\text{\textsc{a}}}X^{\text{\textsc{a}}}
$ with $T_{\text{\textsc{a}}}$ referring to the generators of \textrm{G.}%
\newline
On the boundary, the gauge parameters $X^{\text{\textsc{a}}}$ are 2D scalar
fields $X^{\text{\textsc{a}}}\left( \xi ^{z},\xi ^{\bar{z}}\right) $ and can
be thought of as a bosonic string field carrying an internal charge under G
as it sits into its adjoint representation. Using the generators $T_{\text{%
\textsc{a}}}$ with metric $\mathrm{G}_{\text{\textsc{ab}}}=\left\langle T_{%
\text{\textsc{a}}},T_{\text{\textsc{b}}}\right\rangle $ as well as the
parametrisation:
\begin{equation}
g\left( \xi \right) =e^{T_{\text{\textsc{a}}}X^{\text{\textsc{a}}}\left( \xi
\right) },
\end{equation}%
we can then express $\mathcal{S}_{edge}\left[ g\right] $ in terms of
gradients of $X^{\text{\textsc{a}}}$'s. We first have $g^{-1}\partial
_{z}g=\partial _{z}X$ and similarly $g^{-1}\partial _{\bar{z}}g=\partial _{%
\bar{z}}X;$ they expand respectively like $\sum T_{\text{\textsc{a}}%
}\partial _{z}X^{\text{\textsc{a}}}$ and $\sum T_{\text{\textsc{a}}}\partial
_{\bar{z}}X^{\text{\textsc{a}}}.$ By substituting into $\mathcal{S}_{edge}%
\left[ g\right] $, we end up with the contribution $\int_{\partial
\boldsymbol{M}}\mathrm{G}_{\text{\textsc{ab}}}\partial _{z}X^{\text{\textsc{a%
}}}\partial _{\bar{z}}X^{\text{\textsc{b}}}$ that can be interpreted as the
free field action in the conformal gauge of a bosonic string $X^{\text{%
\textsc{a}}}\left( \xi ^{z},\xi ^{\bar{z}}\right) $ on the group manifold G
with metric $\mathrm{G}_{\text{\textsc{ab}}}$. By using the metric $%
h_{\alpha \beta }$ of the boundary of AdS$_{3}$, $\mathcal{S}_{edge}\left[ g%
\right] $ \textrm{takes the following form:}
\begin{equation}
\int_{\partial \boldsymbol{M}}d^{2}\xi \sqrt{|-h|}h^{\alpha \beta }\mathrm{G}%
_{\text{\textsc{ab}}}\partial _{\alpha }X^{\text{\textsc{a}}}\partial
_{\beta }X^{\text{\textsc{b}}}.
\end{equation}%
Furthermore, \textrm{subsequent to writing} the gauge group elements like $%
g=g_{L}^{-1}g_{R}$ \cite{wbh,wzw3}, the field action (\ref{NCWZW}) for the
non-chiral WZW can be decomposed into two chiral WZW actions: a left block
with field action $\mathcal{S}_{wzw}^{L}\left[ g_{L}\right] $; and a right
one with field action $\mathcal{S}_{wzw}^{R}\left[ g_{R}\right] ;$ their
expressions have similar forms \textrm{to} (\ref{NCWZW}). Focussing on the
boundary terms, the coupling of the left chiral WZW action to $\mathfrak{A}%
_{z}$ reads as \cite{S1,S2}:%
\begin{eqnarray}
\mathcal{S}_{wzw}^{L}\left( g_{L}\right)  &=&\frac{k}{4\pi }\int_{\partial
\boldsymbol{M}}d^{2}zTr\left[ \left( g_{L}^{-1}\partial
_{z}g_{L}^{-1}\right) \left( g_{L}\partial _{\bar{z}}g_{L}\right)
+2g_{L}^{-1}\partial _{\bar{z}}g_{L}\mathfrak{A}_{z}\right]   \notag \\
&&+\frac{k}{12\pi }\int_{\boldsymbol{M}}Tr\left( g_{L}^{-1}dg_{L}\right)
^{3},
\end{eqnarray}%
and the coupling of the right chiral WZW action to $\widetilde{\mathfrak{A}}%
_{\bar{z}}$ is given by a similar term as follows:%
\begin{eqnarray}
\mathcal{S}_{wzw}^{R}\left( g_{R}\right)  &=&\frac{k}{4\pi }\int_{\partial
\boldsymbol{M}}d^{2}zTr\left( g_{R}^{-1}\partial _{z}g_{R}^{-1}g_{R}\partial
_{\bar{z}}g_{R}-2g_{R}^{-1}\partial _{z}g_{R}\widetilde{\mathfrak{A}}_{\bar{z%
}}\right) +  \notag \\
&&\frac{k}{12\pi }\int_{\boldsymbol{M}}Tr\left( g_{R}^{-1}dg_{R}\right) ^{3},
\end{eqnarray}%
Upon variation under gauge transformations, the above chiral field actions
engender two anomalous currents: a right $\mathcal{J}_{\bar{z}}$ and left $%
\widetilde{\mathcal{J}}_{z}$ given by:%
\begin{equation}
\begin{tabular}{lllll}
$\mathcal{J}_{\bar{z}}$ & $=$ & $\frac{k}{2\pi }g_{L}^{-1}\partial _{\bar{z}%
}g_{L}$ & $=$ & $+\frac{k}{2\pi }\mathfrak{A}_{\bar{z}},$ \\
$\widetilde{\mathcal{J}}_{z}$ & $=$ & $-\frac{k}{2\pi }g_{R}^{-1}\partial
_{z}g_{R}$ & $=$ & $-\frac{k}{2\pi }\widetilde{\mathfrak{A}}_{z},$%
\end{tabular}%
\end{equation}%
corresponding to (\ref{anoc}) with a flipped sign exactly as required by the
inflow mechanism. Now let us turn to the gravitational anomaly; as we saw
previously, it is also generated \textrm{by a CS term involving} Christoffel
symbols as:
\begin{equation}
\mathcal{S}_{CS}^{grav}\left( \Gamma \right) =-\frac{c_{L}-c_{R}}{96\pi }%
\int_{\mathcal{M}_{3D}}Tr\left( \Gamma d\Gamma +\frac{2}{3}\Gamma
^{3}\right) ,  \label{CS-grav}
\end{equation}%
with variation given by:
\begin{equation}
\partial _{\mu }T^{\mu \nu }=\frac{c_{R}-c_{L}}{96\pi }g^{\nu \alpha
}\epsilon ^{\mu \rho }\partial _{\beta }\partial _{\mu }\Gamma _{\alpha \rho
}^{\beta }.
\end{equation}%
It has been argued by Carlip \cite{car} that this action as well is
equivalent to a boundary WZW-like action given by:%
\begin{equation}
S_{wzw}^{grav}\left( \left[ \mathrm{\beta }\right] \right) =-\frac{%
c_{R}-c_{L}}{96\pi }\int_{\partial \mathcal{M}_{3D}}\left[ \partial _{z}%
\mathrm{\beta }\partial _{\bar{z}}\mathrm{\beta }+\sqrt{-|h|}\mathrm{\beta }R%
\right] ,  \label{wzw-grav}
\end{equation}%
where $\mathrm{\beta }$ is \textrm{the conjugate of the} vielbein $\mathrm{e}
$ and $R$ is the scalar curvature$.$ The associated variation is therefore:%
\begin{equation}
\partial _{\mu }T^{\mu \nu }=\frac{c_{L}-c_{R}}{96\pi }\epsilon ^{\nu \rho
}\partial _{\rho }R.
\end{equation}%
Taking into consideration the full Polyakov-like (PL) action :%
\begin{eqnarray}
\mathcal{S}_{PL} &=&\int_{\partial AdS_{3}}d^{2}\xi \sqrt{|-h|}h^{\alpha
\beta }\mathrm{G}_{\text{\textsc{ab}}}\partial _{\alpha }X^{\text{\textsc{a}}%
}\partial _{\beta }X^{\text{\textsc{b}}}+  \notag \\
&&\int_{\partial AdS_{3}}i\mathrm{B}_{\text{\textsc{ab}}}\varepsilon
^{\alpha \beta }\partial _{\alpha }X^{\text{\textsc{a}}}\partial _{\beta }X^{%
\text{\textsc{b}}}+  \label{full-poly} \\
&&\int_{\partial AdS_{3}}d^{2}\xi \sqrt{|-h|}h^{\alpha \beta }\partial
_{\alpha }\Phi \partial _{\beta }\Phi +\sqrt{-|h|}\Phi R,  \notag
\end{eqnarray}%
with a \textrm{turned-off} background antisymmetric 2-form tensor $\left(
\mathrm{B}_{\text{\textsc{ab}}}\rightarrow 0\right) ,$ the last term of (\ref%
{full-poly}) corresponds to (\ref{wzw-grav}) with $\Phi =\mathrm{\beta }$.
The full action for our boundary string is therefore:
\begin{equation}
\mathcal{S}_{string}=\mathcal{S}_{wzw}^{L}\left( g_{L}\right) +\mathcal{S}%
_{wzw}^{R}\left( g_{R}\right) +\mathcal{S}_{wzw}^{grav}\left( \mathrm{\beta }%
\right) .
\end{equation}

\subsection{Finiteness of AdS$_{3}$ Landscape}

The Swampland conjecture for the finiteness of the Landscape is often
manifested by an upper bound on the rank of possible gauge groups. This has
been verified for several\textrm{\ }theories in various\textrm{\ }dimensions
\cite{C4, C6, C66}. Here, we test the validity of this Swampland constraint
in AdS$_{3}$/CFT$_{2}$\ and study the implications on the spectrum of higher
spin theories.\textrm{\ } Given the two following: $\left( \mathbf{i}\right)
$ a unitary 2D boundary CFT of AdS$_{3}$ gravity with left central charge $%
c_{L}>0$; and $\left( \mathbf{ii}\right) $ a Kac-Moody algebra $\mathcal{G}%
_{L}$ of level \textrm{k}$_{L}$ realized on the left moving sector of the CFT%
$_{2}$ with central charge as:%
\begin{equation}
0<c_{\mathcal{G}_{L}}=\frac{k_{L}\text{ }\dim \mathcal{G}_{L}}{k_{L}\text{ }%
+h^{\vee }\text{ }},
\end{equation}%
with $h_{L}^{\vee }$ the dual Coxeter number of $\mathcal{G}_{L}$, the
bulk/edge correspondence requires the following consistency constraint
relation:
\begin{equation}
0<c_{\mathcal{G}_{L}}<c_{L}.  \label{ccL}
\end{equation}%
A few comments are in order.\newline
First, regarding the chirality of the current algebras, note that not all
the current algebras are in the left sector. In fact, we started from an AdS$%
_{3}$ action written in a gauge formalism as a difference of two CS actions
accounting for both chiralities left and right as dictated by the AdS$_{3}$
isometry $SO(2,2)=SL(2,\mathbb{R})_{L}\times SL(2,\mathbb{R})_{R}$ (\ref{CS}%
). This bulk symmetry engendered two copies of the W$_{N}$ algebras for the
affine boundary (or two Virasoros for the conformal boundary) with central
charges $c_{L}=c_{R}=3l_{AdS_{3}}/2G$ which became (\ref{CLR}) after
considering the CS gravitational term$.$ The chirality was further set once
we added both gauge boundary terms with CS level $k^{\prime }=-k$ for the
left sector and $\tilde{k}^{\prime }=\tilde{k}$ for the right sector$.$ With
the new found gauge currents we established the equivalence between 't Hooft
anomaly coefficients (\ref{S8}) and the CS levels (\ref{S7}). The sign of 't
Hooft anomaly coefficients determine the chirality of the current algebras
depending on the chosen convention, for instance the right (resp. left)
moving charged currents add positive (resp. negative) contributions to the
anomaly polynomial. By equivalence, this property is transmitted to the CS
levels and therefore we ended up with two copies of the current algebras,
one left because $k^{\prime }=-k<0$ and the other right as $\tilde{k}%
^{\prime }=\tilde{k}>0$. Taking advantage of this chirality, it suffices to
compute the consistency constraint (\ref{ccL}) on a particular sector, left
one in this case $k=k_{L}$, to derive the bound on the rank.\newline
Second, concerning the origin of $c_{G};$ since now both boundary gauge
components contribute to the boundary charge $Q_{{\small \zeta }}=Q_{{\small %
\zeta }}^{(z)}+Q_{{\small \zeta }}^{(\bar{z})}$ with:%
\begin{equation}
Q_{{\small \zeta }}^{(z)}=-\frac{k}{2\pi }\int tr\left( \zeta \mathfrak{A}%
_{z}\right) d\xi ^{z}\qquad ,\qquad Q_{{\small \zeta }}^{(\bar{z})}=-\frac{k%
}{2\pi }\int tr\left( \zeta \mathfrak{A}_{\bar{z}}\right) d\xi ^{\bar{z}},
\end{equation}%
we therefore get in addition to the usual asymptotic algebra (\ref{KM})
given by the charges of $\mathfrak{A}_{-}$, a second algebra (\ref{ex1})
given by the chemical potentials of $\mathfrak{A}_{+}.$ First, notice that
central charges in the inequality (\ref{ccL}) namely $c_{G}\leq c_{L}$
concern the CFT$_{2}$ living on the boundary of AdS$_{3}.$\ The $c_{L}$ is
the usual anomaly charge in the left sector of the CFT$_{2};$ the anomaly
charge for the right sector is denoted $c_{R}.$ The $c_{G}$ is a conformal
anomaly induced by \textit{our generalization of GR boundary condition}
given by $\delta \mathfrak{A}_{+}=0$ while $\delta \mathfrak{A}_{-}\neq 0.$
\emph{For the GR boundary condition}\textit{, }there is no $c_{G},$ and as
such \emph{eq(\ref{ccL}) reduces} to $c_{L}\geq 0$ which is just the
unitarity condition of the CFT.\newline
By generalizing the GR boundary condition to $\delta \mathfrak{A}_{+}\neq 0$
and $\delta \mathfrak{A}_{-}\neq 0,$\ one obtains, besides the $c_{L}$
anomaly symmetry, an additional asymptotic current algebra $G_{\mathrm{k}}$
with Kac-Moody level $k,$ and central conformal charge anomaly $c_{G}$ given
by the Suggawara construction.\newline
\textrm{This generalisation of the }GR boundary condition\textrm{\ relies on
treating} $\delta \mathfrak{A}_{+}$ \textrm{and} $\delta \mathfrak{A}_{-}$
\textrm{in a somehow equal footing; with a cost involving the emergence of a
non trivial }$c_{G}.$ The obtained boundary CFT is different from the GR
based conformal theory; this is a natural behaviour because the boundary
condition has been modified from GR towards extended GR. The novelty is that
the new behaviour is manifested by the shift from $c_{L}\geq 0$ (for GR
boundary) to $c_{L}-c_{G}\geq 0$ (for extended GR boundary); the minus sign
has been interpreted in terms of reaction to the change of the boundary
conditions due to the presence of strings as proposed in the paper; see
subsubsec \textbf{4.2.2}.\newline
Third, regarding the derivation of the condition $c_{L}\geq c_{G}$ in
connection with the Swampland program, it has been interpreted in terms of
the weak gravity conjecture that requires the gravitational force to be the
weakest in any effective field theory coupled to gravity. Following \textrm{%
Brown \& Hennaux} \cite{A2}, $c_{L}$ is related to the Newton constant $G_{N}
$ and the AdS$_{3}$ radius as $c_{L}=3l_{AdS_{3}}/2G_{N};$ as such it can be
naturally interpreted as a measure of the "gravitational strength" \textrm{%
\cite{wgc3}} so that \textrm{a weak gravity coupling corresponds to a
greater central charge compared to the one associated with gauge symmetry}.
\newline
The formula (\ref{ccL}) therefore insures that gravity is the weakest force
within the theory by demanding c$_{L}$ to be the greater central charge
corresponding thus to a weak gravity coupling as shown by (\ref{CLR}) given
in the vanishing limit of $\mathrm{\beta }$ by:
\begin{equation}
G_{N}=\frac{3l}{2c}.
\end{equation}%
Within this view, the\textrm{\ }above constraint (\ref{ccL}) is nothing but
the weak gravity conjecture that we state as:%
\begin{equation}
\frac{3l}{2c_{L}}<\frac{3l}{2c_{\mathcal{G}_{L}}}<\infty .
\end{equation}%
Applying this description to a left Kac-Moody algebra $\mathcal{G}_{k_{L}}$
with level $\mathrm{k}_{L}$ given by the special linear family $sl(N)_{k_{L}}
$, the Swampland constraint (\ref{ccL}) reads like $c_{sl(N)_{L}}\leq c_{L}$%
; thus leading to a constraint relation on the integers \textrm{k} and N
namely:%
\begin{equation}
0<\frac{k_{L}(N^{2}-1)}{k_{L}+N}\leq c_{L}.  \label{B4}
\end{equation}%
We have already discussed the coupling of spin-s fields to AdS$_{3}$ gravity
in the three dimensional Chern-Simons formulation based on the split real
form $sl\left( N\right) _{L}\times sl\left( N\right) _{R}$ where each spin $%
s=2...N$ appears once meaning that $2\leq s\leq N.$ Focusing on the case of
one $sl(N)_{L}$\ factor with (\ref{B4}) and $k_{L}=1$, we get:%
\begin{equation}
s\leq N\leq c_{L}+1.  \label{B5}
\end{equation}%
Following the same description to left Kac-Moody algebra $\mathcal{G}_{k}^{L}
$ given by orthogonal and symplectic gauge symmetries with Kac-Moody level $%
k_{L}=1$, the swampland constraint (\ref{ccL}) reads for the corresponding
higher spin families as follows:%
\begin{eqnarray}
SO(N,1+N)_{L} &:&\qquad \frac{N(2N+1)}{1+2N-1}\leq c_{L},  \label{B61} \\
SO(N,N)_{L} &:&\qquad \frac{N(2N-1)}{1+2N-2}\leq c_{L},  \label{B62} \\
Sp\left( 2N,\mathbb{R}\right) _{L} &:&\qquad \frac{N(2N+1)}{1+N+1}\leq c_{L},
\label{B63}
\end{eqnarray}%
where we have substituted $h_{\text{\textsc{so}}_{N}}^{\vee }=N-2$ and $h_{%
\text{\textsc{sp}}_{2N}}^{\vee }=N+1$. This leads to the following upper
bounds:%
\begin{eqnarray}
SO(N,1+N)_{L} &:&\qquad N\leq c_{L}-\frac{1}{2}, \\
SO(N,N)_{L} &:&\qquad N\leq c_{L}, \\
Sp\left( 2N,\mathbb{R}\right) _{L} &:&\qquad N\leq \frac{1}{4}\left(
c_{L}-1\right) +\frac{1}{4}\sqrt{c_{L}^{2}+14c_{L}+1}.
\end{eqnarray}%
Next, we will compare the finiteness constraint with the gravitational
exclusion principle derived from a BTZ black hole consideration for the $%
sl(N)$ case and excerpt possible constraints on the CS level \textrm{k}. But
before we will refine the bound using the AdS distance conjecture.

\subsection{The AdS distance conjecture and the finiteness of the AdS$_{%
\emph{3}}$ Landscape}

Outwardly by juxtaposing the AdS distance conjecture (ADC) and the Landscape
finiteness constraint, both statements look conflicted and the finiteness
appears to be violated by the AdS space in the absence of a cut-off scale.
The AdS$_{3}$ distance conjecture stipulates the existence of an infinite
tower of massless states upon taking $\Lambda =-1/l_{AdS_{3}}^{2}\rightarrow
0$ implying that $l_{AdS_{3}}\rightarrow \infty ,$ given a mass scale $m\sim
$ $|\Lambda |^{\alpha }$ with $\Lambda $ being the cosmological constant of
the AdS$_{3}$ space and $\alpha $\ \textrm{a positive constant} \cite{adswp}%
. Accordingly, in order to ensure the non-violation of the finiteness
principle we need to impose a cut-off for our AdS$_{3}$ as follows \cite{rev}
\begin{equation}
\Lambda _{cut-off}\leq m\sim |\Lambda |^{\alpha }\sim 1/l_{AdS_{3}}^{2\alpha
}\qquad \text{where }l_{AdS_{3}}\text{ is fixed.}
\end{equation}%
The consideration of the cut-off $\Lambda _{cut-off}$ further sharpens the
finiteness conjecture. In fact by fixing the cut-off, the number of AdS
theories\ and the associated number of the dual CFTs become finite. To see
this, let's rewrite the central charge in terms of the cut-off like:%
\begin{equation}
\Lambda \sim c^{\frac{-2}{d-2}}\sim c^{-2},
\end{equation}%
with%
\begin{equation}
\Lambda _{cut-off}\leq m\sim |\Lambda |^{\alpha }\sim \frac{1}{%
l_{AdS_{3}}^{2\alpha }},
\end{equation}%
we get:
\begin{equation}
c\leq \Lambda _{cut-off}^{-\frac{1}{2\alpha }},
\end{equation}%
and therefore the number of the CFT theories and their holographic dual is
finite. Furthermore, the rank of family of solutions corresponding to the HS
gauge symmetry is also upper bounded by the cut-off since:%
\begin{equation}
c_{G}\left( N\right) \leq c\leq \Lambda _{cut-off}^{-\frac{1}{2\alpha }}.
\label{finit}
\end{equation}%
On top of this, for higher spin gravity, a constraint on the rank of the
higher spin symmetry implies a restraint on the highest allowed spin within
the theory, for the linear case for instance it reads:%
\begin{equation}
\begin{tabular}{|c|c|c|}
\hline
symmetry & rank & $s\leq f\left( rank_{G}\right) $ \\ \hline
$sl(N,\mathbb{R})$ & $N$ & $s\leq N$ \\ \hline
\end{tabular}
\label{finit2}
\end{equation}%
In conclusion, the finiteness conjecture for higher spin gravity in AdS$_{3}$
is threefolded. First, given a cut-off $\Lambda _{cut-off}$ the sum of all
AdS spaces and the corresponding conformal theories is finite which is
implied by the right side of the inequality (\ref{finit}). Second, the gauge
symmetries must have a finite rank which is implied by the left side of (\ref%
{finit}). And lastly, the finiteness of the spin spectrum implied by (\ref%
{finit2}). Our definition of the AdS$_{3}$ Landscape in subsection \textbf{%
4.1} is therefore sharpened and now comprises theories verifying all of (\ref%
{finit}) and (\ref{finit2}), the AdS$_{3}$ Landscape is thus a finite set of
AdS$_{3}$ theories with a non vanishing cosmological constant $\Lambda $\
and their holographic duals with central charges below the cut-off $\Lambda
_{cut-off}$. The higher spin gauge symmetries of the AdS$_{3}$ gravity have
a finite rank and an upper bound of the highest allowed spin. In contrast,
any theory in subsection \textbf{4.1} violating either (\ref{finit}) or (\ref%
{finit2}) is considered a Swampland theory. Nonetheless, there is an
exception for certain theories defined independently of a cut-off $\Lambda
_{cut-off}$ \ where the massless modes are localised in a defect and the
corresponding infinite families are symmetries of these defects and since
there are no Swampland constraints that forbid from having infinite massless
modes localised in a defect, these theories are consistent and don't violate
the ADC nor the finiteness conjecture as detailed in \cite{adswp},\cite{rev}.

\subsection{Finiteness conjecture versus gravitational exclusion principle}

As discussed in the subsection \textbf{4.4}, the finiteness conjecture in a
quantum AdS$_{3}$ gravity theory is a consistency condition that must be
verified by 3D Landscape models. For the BTZ black hole, this fundamental
constraint has been derived and known as the gravitational exclusion
principle (GEP). It is simply stated as the requirement for the number of
higher spin version of boundary gravitons to obey the upper bound appointed
by Cardy's formula. The gravitational exclusion principle in the AdS$_{3}$
space was derived in \cite{G2} by studying the partition function of the BTZ
black hole before and after considering higher spin fields. For an Euclidean
BTZ black hole with 2-torus topology at its asymptote, the partition
function is given by:%
\begin{equation}
Z_{BTZ}\left( \tau ,\bar{\tau}\right) =Tr\left( q^{L_{0}}\bar{q}^{\bar{L}%
_{0}}\right) =\sum_{\Delta ,\bar{\Delta}}d\left( \Delta ,\bar{\Delta}\right)
q^{\Delta }\bar{q}^{\bar{\Delta}},  \label{BH1}
\end{equation}%
where $\Delta ,\bar{\Delta}$ are conformal weights equivalent to $-c/24$ for
the ground state (for a standard CFT normalization) and $d\left( \Delta ,%
\bar{\Delta}\right) $ is the entropy or number of states given by Cardy's
formula:%
\begin{equation}
\log \left( d\left( \Delta ,\bar{\Delta}\right) \right) \sim 2\pi \sqrt{%
\frac{c\Delta }{6}}+2\pi \sqrt{\frac{c\bar{\Delta}}{6}}.  \label{BH2}
\end{equation}%
The bittersweet of Cardy's formula is that we don't exactly know what really
contributes or don't to the entropy as we didn't need the full states
spectrum to construct it, we just took advantage of the presence of a CFT at
the asymptotic limit of AdS$_{3}$ and yield the density of its states (\ref%
{BH2}). As for the higher spin fields, we don't have the exact formulation
of the partition function $Z\left( q\right) $, but we do have the tree and
the one loop contributions. The full spectrum is given by a set of states,
constructed by acting on the vacuum state $Z\left( q\right) ^{\left(
0\right) }$ by the vacuum character $\chi _{N}\left( q\right) $ of the W$_{N}
$ algebra:
\begin{equation}
\begin{tabular}{lll}
$Z\left( q\right) ^{\left( 0\right) }$ & $=$ & $q^{-c/24}\bar{q}^{-c/24},$
\\
$Z\left( q\right) ^{\left( 1\right) }$ & $=$ & $q^{-c/24}\bar{q}%
^{-c/24}|\chi _{N}\left( q\right) |^{2},$%
\end{tabular}
\label{BH3}
\end{equation}%
with $q=e^{(2\pi i\tau )}$ and $\tau $ is the conformal structure of the
2-torus at the boundary of AdS$_{3}$ space. For a finite N, we have \textrm{%
\cite{jeh, jeh1,jeh2}}:%
\begin{equation}
\chi _{N}\left( q\right) =\prod_{s=2}^{N}\prod_{n=s}^{\infty }\left(
1-q^{n}\right) ^{-1}=\sum_{\Delta }p_{\Delta }^{N}q^{\Delta },  \label{BH4}
\end{equation}%
where the coefficients $p_{\Delta }^{N}$ can be derived for a large $\Delta $
as%
\begin{equation}
\log \left( p_{\Delta }^{N}\right) \simeq 2\pi \sqrt{\frac{\left( N-1\right)
\Delta }{6}}.  \label{BH5}
\end{equation}%
We can see in (\ref{BH4}) how the number of states\ escalates with the
addition of the higher spin fields as given here below for the two leading
values:
\begin{equation}
N=2:\qquad \chi _{2}\left( q\right) =\prod_{s=2}^{2}\prod_{n=s}^{\infty
}\left( 1-q^{n}\right) ^{-1}=\prod_{n=2}^{\infty }\left( 1-q^{n}\right)
^{-1},  \label{BH56}
\end{equation}%
and%
\begin{equation}
N=3:\qquad \chi _{3}\left( q\right) =\prod_{s=2}^{3}\prod_{n=s}^{\infty
}\left( 1-q^{n}\right) ^{-1}=\prod_{n=2}^{\infty }\left( 1-q^{n}\right)
^{-1}\times \prod_{n=3}^{\infty }\left( 1-q^{n}\right) ^{-1}.
\end{equation}%
But in (\ref{BH2}) Cardy's formula restrained the number of states, since
the black hole has the greater entropy. Thus, in order for (\ref{BH5}) to
describe states from the spectrum of the theory we must have:%
\begin{equation}
N-1\leq c.  \label{BH6}
\end{equation}%
In the following, we will pinpoint some important implications of this
constraint: $\left( \mathbf{i}\right) $ A remarkable aspect of the AdS$_{3}$
gravity is that it implicates BTZ black hole solutions, where the asymptotic
density of states (\ref{BH2}) is equivalent to their Bekenstein-Hawking
entropy:%
\begin{equation}
S_{entropy}=\frac{2\pi r_{+}}{4G},\qquad r_{+}=2\sqrt{Gl_{AdS_{3}}}\left(
\sqrt{L_{0}}+\sqrt{\bar{L}_{0}}\right) .
\end{equation}%
Hence, (\ref{BH6}) shows that the BTZ black hole states form an upper bound
on the spectrum of the AdS$_{3}$ gravity. $\left( \mathbf{ii}\right) $ This
constraint is only guaranteed for non perturbative theories; for theories
with values of $N$ and $c$ that violate the constraint, either some of the
perturbative states will be removed from the spectrum as a result of a
linearization instability within the theory or the value of $%
G_{N}/l_{AdS_{3}}$ will be renormalized so that the constraint stays
verified. $\left( \mathbf{iii}\right) $ The gravitational exclusion
principle formula in (\ref{BH6}) was derived following consideration from
the asymptotic border symmetry W$_{N},$ what is interesting is that it has
the same implications as the one we derived from the bulk symmetry $sl\left(
N\right) $ based on the 3D Swampland conjecture (\ref{B4}):%
\begin{equation}
N-1\leq c\qquad \iff \qquad \frac{k(N^{2}-1)}{k+N}\leq c.  \label{res}
\end{equation}%
As for the possible values of the level $k=k_{L},$ it has been conjectured
in \cite{witten} and as stated in above (\ref{level}), the CS levels are
positive integers. The case where $k=k_{L}=0$ is proper to trivial global
symmetries \cite{S1}, it will lead to the vanishment of the currents with an
inconsequential variational principal. The next possible value would be $%
k=k_{L}=1,$ which was used in subsection \textbf{4.3} as the first possible
value to compute the bound, and a priori there is no reason whatsoever from
the anomaly inflow argument to prevent any other positive integer value. In
fact, $k=k_{L}\varpropto $ $l_{AdS_{3}}/4G$ and since we have already
established that the only requirement for $l_{AdS_{3}}$ is that it should be
fixed and finite, the level can therefore run several different values. But
the comparison (\ref{res}) shows that at this particular level $k=k_{L}=1,$
with $l_{AdS_{3}}=4G$ describing a semi classical regime where $%
l_{AdS_{3}}>>G,$ both the swampland constraint derived from the anomaly
consideration and the gravitational exclusion principal derived from a BTZ
black consideration before and after considering high spin fields are
equivalent and lead to the same upper bound on the rank. Conclusively, the
adopted value of the level $k=k_{L}=1$ isn't a requirement from the anomaly
analysis but a requisite from juxtaposing both constraints.\newline
Furthermore in \cite{maloney}, the authors suggested that the generalization
of the bound $N-1\leq c$ to other algebras can be achieved by replacing $N-1$
with the central charge of the associated current algebra:
\begin{equation}
c_{current}=\sum_{i}\frac{k_{i}\text{ }\dim \mathcal{G}_{i}}{%
k_{i}+h_{i}^{\vee }\text{ }}.
\end{equation}%
Which means that our bound derived in the previous section (\ref{BH6}) for
the rest of the higher spin families still holds for this generalisation,
and we will have a similar correspondence just as (\ref{res}) for each
considered algebra. For the orthogonal class, we have:%
\begin{eqnarray}
SO(N,1+N) &:&\quad \log d\left( \Delta ,\bar{\Delta}\right) \sim 2\pi \sqrt{%
\frac{N+\frac{1}{2}}{6}}\left( \sqrt{\Delta }+\sqrt{\bar{\Delta}}\right) , \\
SO(N,N) &:&\quad \log d\left( \Delta ,\bar{\Delta}\right) \sim 2\pi \sqrt{%
\frac{N}{6}}\left( \sqrt{\Delta }+\sqrt{\bar{\Delta}}\right) .
\end{eqnarray}

\section{Conclusion and comments}

\label{sec5} Throughout this work, we aimed to assess the three dimensional
topological gravity from the perspective of Swampland conjectures
particularly the finiteness constrain. At first, we presented the
Chern-Simons formulation of the three dimensional anti de Sitter gravity
with the $sl\left( 2,\mathbb{R}\right) $ algebra. Then, we proceeded by
deriving the extended asymptotic symmetry algebra and showed the holographic
duality between the bulk AdS$_{3}$ and the boundary CFT$_{2}$. When
generalising the $sl(2)$ algebra to describe higher spin theories via the
principal embedding of $sl(2)$, we proposed a classification of these new
symmetries using split real forms of Lie algebras to form the AdS$_{3}$
Landscape. We later checked for the validity of the finiteness constraint
for topological gravity in the three dimensional case by deriving the bound
on the rank of Landscape gauge symmetries and studied the implications on
the highest spin of HS theories by taking advantage of the connection
between the spin and the rank of the theory. Finally, we intersected the
finiteness conjecture with the AdS distance conjecture to further refine the
bound and provided further arguments from Literature by comparing the
Swampland constraint with the gravitational exclusion principle and showed
how the correspondence between these two bounds lead to constrain the CS
level\textrm{\ k}.\newline
An interesting inquiry one can ponder upon next, is how the finiteness
conjecture tackles the infinite higher spin theories. These are AdS$_{3}$
higher spin theories formulated via Chern-Simons actions each based on a
copy of a one-parameter Lie algebra $hs[\mu ]\oplus hs[\mu ]$ such that for $%
\mu =N$, this symmetry reduces to the regular $sl(N)\oplus \ sl(N)$ with a
finite number of higher spin gauge fields \cite{ihs1}. With this
description, one can envision these theories as an extension of the finite $%
sl(N)$ models for non-integer $N$.\newline
As proved in \cite{ihs2, ihs3}, these bulk AdS$_{3}$ theories are dual to
the $W_{\infty }[\mu ]$ algebra with conformal spin currents $W^{\left(
s\right) }$ where $s$ is unbounded, meaning that $s=2,3,..$. And similarly
to the bulk reduction, for $\mu =N$ the $W_{\infty }[\mu ]$ algebra is
reduced to the $W_{N}$ algebra, the conformal dual to the $sl(N)$ algebra.
Furthermore in \cite{ihs1}, it was shown that the $W_{\infty }[\mu ]$
algebra is isomorphic to the $W_{N,k}$ minimal models such that:%
\begin{equation}
W_{N,k}\cong W_{\infty }\left[ \lambda \right] ,\qquad \text{\ with 't Hooft
coupling }\lambda =\frac{N}{N+k},
\end{equation}%
where the $W_{N,k}$ minimal models are given by the coset:%
\begin{equation}
\frac{G_{k}\otimes G_{1}}{G_{k+1}},
\end{equation}%
for the usual $G=SU\left( N\right) $, it becomes:
\begin{equation}
\frac{SU\left( N\right) _{k}\otimes SU\left( N\right) _{1}}{SU\left(
N\right) _{k+1}},
\end{equation}%
with central charge:%
\begin{equation}
c_{N,k}=\left( N-1\right) \left[ 1-\frac{N\left( N-1\right) }{\left(
N+k\right) \left( N+k+1\right) }\right] .  \label{cnk}
\end{equation}%
With this isomorphism, it was argued in \cite{ihs4} that the infinite higher
spin theory $hs[\mu =\lambda ]$ in AdS$_{3}$ is dual to the conformal $%
W_{N,k}$ minimal models with a 't Hooft limit given as follows:
\begin{equation}
0\leq \lambda =\frac{N}{N+k}\leq 1\qquad \text{for }N,k\rightarrow \infty .
\end{equation}%
Moreover, the full spectrum of the conformal $W_{N,k}$ minimal models
requires the addition of two complex scalar fields. The bulk therefore
becomes an AdS$_{3}$ theory with an infinite higher spin symmetry $hs[\mu
=\lambda ]\oplus hs[\mu =\lambda ],$ coupled to two complex scalar fields.
As we can see from (\ref{cnk}), the central charge for the minimal models
verify:%
\begin{equation}
c_{N,k}=\left( N-1\right) \left[ 1-\frac{N\left( N-1\right) }{\left(
N+k\right) \left( N+k+1\right) }\right] \leq \left( N-1\right) ,
\label{nfinit}
\end{equation}%
which clearly violate the left side of the bound (\ref{finit}) and seems a
priori non-consistent and belong to the Swampland. However the holographic
principle as stated in \cite{gep2}, constrains the number of degrees of
freedom in a space region proportionally to its area in agreement with the
GEP in AdS$_{3}$. Which means that certain excitation states should be
removed to accommodate the finiteness of the area of the space region, for
the reason that these aren't in fact linearizations of true states in the
Hilbert space; in this case the theory is said to have a linearization
instability.\newline
The mechanism with which we remove the precise number of states needed in
order to saturate both the bound (\ref{finit}) and (\ref{nfinit}), resulting
in a maximum number of states given by the central charge $c$, is thoroughly
described in \cite{gep2}. The authors showed that the norm of the apparent
innocuous states of the $W_{N,k}$ minimal models is zero, hence the theory
has null vectors that should not be considered as physical states and in
consequence should be removed from the spectrum.\newline
From the perspective of the Swampland program, the bound is further
sharpened. In the 't Hooft limit the central charge behaves as $c\rightarrow
\infty $ and the AdS$_{3}$ radius $l_{AdS_{3}}$ becomes infinite, in
adjustment with the infinite number of states as dictated by the holographic
principle, which we forbid in accordance with the AdS$_{3}$ distance
conjecture. Consequently in order for the theories to be considered in the
AdS$_{3}$ Landscape, the right side of (\ref{finit}) should also be
verified; $c,$ as well as the corresponding radius $l_{AdS_{3}},$ can grow
big but never infinite so that the number of CFTs is finite as required by
the finiteness conjecture in AdS.\newline
As a continuation of this work, we would like to check for the validity of
the bound on the rank when the theory is supersymmetrized, especially for
the three dimensional supergravity case based on superalgebras. \appendix

\section{Appendix: Graphical description of the principal embedding}

\label{app}\textbf{\ }In this appendix, we give some aspects regarding the
construction of higher spin AdS$_{3}$ gravity by using graphical tools. We
illustrate this construction on \textrm{two particular models: the first has
an} $SL(5,\mathbb{R})$ gauge symmetry; it is a representative example of the
linear $SL(N,\mathbb{R})$ family. The second \textrm{has an }$SO(5,5)$
symmetry; it is a representative example of the real split form of the
complex orthogonal $\boldsymbol{D}_{N}$ family.\newline
We begin by recalling that as for Dynkin graphs of finite Lie algebras, to
each node $\mathfrak{n}_{i}$ ($1\leq i\leq N$) of a given Tits-Satake
diagram, one has a simple root $\alpha _{i}$ and then a $sl(2,\mathbb{R})$
Lie algebra which is isomorphic to $so(1,2).$ This graphic feature shows
that $sl(2,\mathbb{R})$ and $so(1,2)$ are the building blocks of $sl(5,%
\mathbb{R})$ and $so(5,5).$ The $sl(5,\mathbb{R})$ involves four $sl(2,%
\mathbb{R})$s and the $so(5,5)$ has five $so(1,2)$s. The higher spin
construction fixes one of these $sl(2,\mathbb{R})$ [resp. $so(1,2)]$ as a
fundamental symmetry; and the generators of the coset space $sl(5,\mathbb{R}%
)/sl(2,\mathbb{R})$ [resp. $so(5,5)/so(1,2)]$ are imagined in terms of
higher dimensional representations of $sl(2,\mathbb{R})$ [resp. $so(1,2)]$.%
\newline
In our approach, the fundamental $sl(2,\mathbb{R})$ [resp. $so(1,2)]$ is
taken as an extremal node of the Tits-Satake diagram; say the most left node
$\mathfrak{n}_{1}$ or the most right node $\mathfrak{n}_{N}.$ For the most
right node, the fundamental $sl(2,\mathbb{R})$ [resp. $so(1,2)]$ corresponds
to the simple root $\alpha _{1}$ and is generated by the Cartan-Weyl
operators $\{H_{\alpha _{1}},E_{\pm \alpha _{1}}\}$ which are equivalent to
the usual Cartesian $\{J_{a}\}\equiv \{J_{0},J_{1},J_{2}\}.$

\begin{figure}[tbph]
\begin{center}
\includegraphics[width=6cm]{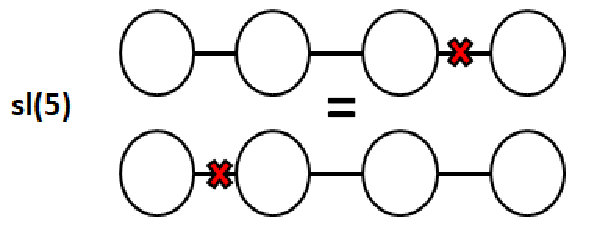}
\end{center}
\par
\vspace{-0.5cm}
\caption{Tits-Satake diagram for the split real form of the linear model $%
SL(5,\mathbb{R})$.}
\label{F1}
\end{figure}

$\bullet $ \emph{HS model with} $sl(5,\mathbb{R})$\newline
For this particular model, the principal embedding of $sl(2,\mathbb{R})$
into the $sl(5,\mathbb{R})$ Lie algebra can be nicely described by
Tits-Satake diagrams. The $sl(5,\mathbb{R})$ is contrived by the cutting of
an extremal node in its Tits-Satake diagram; the cut node $\alpha _{1}$
defines the fundamental $sl(2,\mathbb{R})$. Higher spin operators are given
by the generators of the coset space $G/H$ with $G=SL(5,\mathbb{R})$ and $%
H=SL(2,\mathbb{R}).$ Notice that for the $sl(N,\mathbb{R})$ series, the left
and the right node cut lead to identical decompositions. This is because
both most left and most right cuttings give the Tits-Satake diagram\ of $%
sl(N-1,\mathbb{R})$ as depicted by the \textrm{figure }\ref{F1} for the
example $sl(5,\mathbb{R})\rightarrow sl(4,\mathbb{R})$.
In this model, the 24 generators of $sl(5,\mathbb{R})$ splits as $3+5+7+9$
with the first 3 referring to the three generators $\{J_{0},J_{1},J_{2}\}$
of the basic $sl(2,\mathbb{R})$; and the other odd $2\mathfrak{j}+1$ giving
the remaining generators of $sl(5,\mathbb{R})$ sitting in the irreducible
isospin $\mathfrak{j}$ representations of $sl(2,\mathbb{R})$. For the $sl(5,%
\mathbb{R})$ model, we only get one higher spin fields spectrum with
boundary conformal spin ranging in $s=2,3,4,5.$ This is because the most
left cutting node and the most right cutting one are equivalent. \newline
The AdS$_{3}$/CFT$_{2}$ theory therefore becomes a bulk $sl(5,\mathbb{R}%
)\oplus sl(5,\mathbb{R})$ higher spin AdS$_{3}$ gravity theory with a 2D
boundary CFT where the asymptotic symmetry algebra is given by two copies of
the $\boldsymbol{W}_{sl\left( 5,\mathbb{R}\right) }$-algebra, $\boldsymbol{W}%
_{sl\left( 5,\mathbb{R}\right) }\oplus \boldsymbol{W}_{sl\left( 5,\mathbb{R}%
\right) }$. The higher spin partition function of the CS realisation of HS
AdS$_{3}$ gravity with gauge symmetry $G=SL\left( 5,\mathbb{R}\right) $
reads as:%
\begin{equation}
\mathcal{Z}_{G}=|\mathbf{\chi }_{G}\left( q\right) |^{2},
\end{equation}%
with:
\begin{equation}
\mathbf{\chi }_{G}=\frac{1}{\left[ \mathbf{\eta }\left( q\right) \right] ^{4}%
}q^{-\frac{c-4}{24}}\left( 1-q\right) ^{4}\left( 1-q^{2}\right) ^{3}\left(
1-q^{3}\right) ^{2}\left( 1-q^{4}\right) ,
\end{equation}%
where $q$ is the usual $e^{i2\pi \tau }$ of the 2-torus, and $\mathbf{\eta }%
\left( q\right) $ is the Dedekind function. Moreover, knowing that for the
fundamental symmetry $H=SL\left( 2,\mathbb{R}\right) $ the vacuum character $%
\ \mathbf{\chi }_{H}$ is equal $\frac{1}{\left[ \mathbf{\eta }\left(
q\right) \right] }q^{-\frac{c-1}{24}}\left( 1-q\right) ,$ it results that
the partition function for conformal spins beyond $s=2$ is given by:%
\begin{equation}
\mathbf{\chi }_{G/H}=\frac{1}{\left[ \mathbf{\eta }\left( q\right) \right]
^{3}}q^{\frac{1}{8}}\left( 1-q\right) ^{3}\left( 1-q^{2}\right) ^{3}\left(
1-q^{3}\right) ^{2}\left( 1-q^{4}\right) .
\end{equation}

\begin{figure}[tbph]
\begin{center}
\includegraphics[width=5cm]{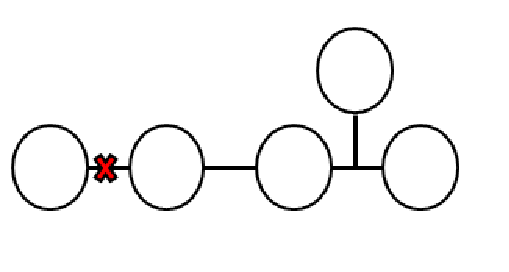}
\end{center}
\par
\vspace{-0.5cm}
\caption{Left node decomposition of Tits-Satake diagram for the split real
form of B$_{5}$.}
\label{Fig3}
\end{figure}

\begin{figure}[tbph]
\begin{center}
\includegraphics[width=5cm]{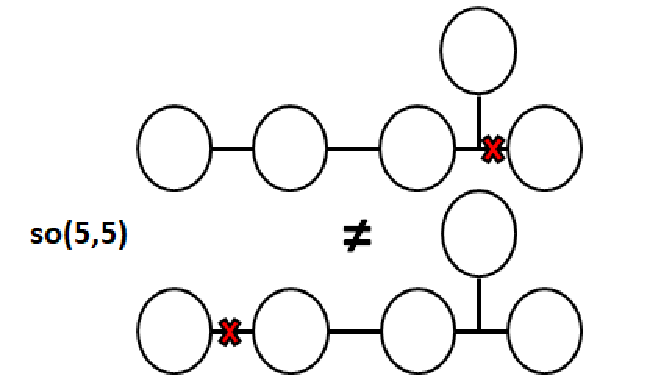}
\end{center}
\par
\vspace{-0.5cm}
\caption{Tits-Satake diagram for the split real form of the linear family D$%
_{5}$.}
\label{Fig4}
\end{figure}

$\bullet $ \emph{HS model with} $so(5,5)$\newline
Contrary to $sl(5,\mathbb{R})$, the extremal node decomposition of the
Tits-Satake diagram of the $so(5,5)$ symmetry gives two different spectrums
that can be termed as: $\left( \mathbf{i}\right) $ \emph{vectorial} with HS
spectrum ranging in $s=\left\{ 2,3,4;7,9\right\} $; and $\left( \mathbf{ii}%
\right) $ \emph{spinorial} with spin range $s=\left\{ 2,3,4;5,11\right\} $.
The vectorial case corresponds to cutting the left node in the $so(5,5)$
Tits-Satake diagram; see Figure \ref{Fig3}.
For this case we have the branching $so(5,5)\rightarrow so(4,4)$. The 45
dimensions of $so(5,5)$ decomposes into irreducible representations of the
fundamental $so(1,2)$ namely $45=3+5+7+13+17.$ The first 3 are associated
with the simple root $\alpha _{1}$; they give the three generators of $%
so(1,2)$. The other odd $2\mathfrak{j}+1$ give the remaining generators of $%
sl(5,\mathbb{R})$; they sit in the irreducible $so(1,2)$-isospins $\mathfrak{%
j}=1,2,3;6,8.$ As for the partition function of this vectorial model with
symmetry $G=SO(5,5)$ and $H=SO(1,2)$, it is given by:
\begin{equation}
\mathcal{Z}_{G}^{vect}=|\mathbf{\chi }_{G}^{vect}|^{2},
\end{equation}%
with%
\begin{equation}
\mathbf{\chi }_{G}^{vect}=\frac{q^{-\frac{c}{24}}}{[\mathbf{\eta }\left(
q\right) ]^{5}}q^{\frac{5}{24}}\frac{\left( 1-q^{7}\right) \left(
1-q^{8}\right) }{\left( 1-q\right) \left( 1-q^{2}\right) }%
\prod\limits_{n=1}^{8}\left( 1-q^{n}\right) ^{7-n}.
\end{equation}%
For the spinorial case, the allowed higher spins are obtained by cutting the
most right node $\mathfrak{n}_{5}$ in the Tits-Satake diagram of $so(5,5)$.
Because of the Z$_{2}$- outer-automorphism of the $so(5,5)$ diagram, one can
also cut the $\mathfrak{n}_{4}$ without affecting the final result as shown
by \textrm{Figure }\ref{Fig4}.
Here, we have the branching $so(5,5)\rightarrow sl(4,\mathbb{R})$ and the 45
dimensions decomposes as $45=3+5+7+9+21$; the blocks in this splitting
correspond to the $so(1,2)$-isospins $\mathfrak{j}=1,2,3,4,10.$ As for the
partition function of this vectorial model with symmetry $G=SO(5,5)$ and $%
H=SO(1,2)$, it is given by:
\begin{equation}
\mathcal{Z}_{G}^{spin}=|\mathbf{\chi }_{G}^{spin}|^{2},
\end{equation}%
with%
\begin{equation}
\mathbf{\chi }_{G}^{spin}=\frac{q^{-\frac{c}{24}}}{\left[ \mathbf{\eta }%
\left( q\right) \right] ^{5}}q^{\frac{5}{24}}\frac{\left( 1-q^{7}\right)
\left( 1-q^{8}\right) }{\left( 1-q\right) \left( 1-q^{2}\right) }%
\prod\limits_{n=1}^{8}\left( 1-q^{n}\right) ^{7-n}.
\end{equation}%
\begin{equation*}
\end{equation*}

\end{document}